\documentclass[sigconf]{acmart}
\usepackage{graphicx}
\usepackage{times}
\usepackage{pbox}
\usepackage{footnote}
\makesavenoteenv{tabular}
\makesavenoteenv{table}
\usepackage{mathtools}
\usepackage{url}
\usepackage{balance}  
\usepackage{enumitem}
\usepackage{array}
\usepackage{booktabs}
\newtheorem{mydef}{Definition}
\newtheorem{mylemma}{Lemma}

\usepackage{pifont}

\usepackage[ruled,linesnumbered]{algorithm2e}
\usepackage{tikz}

\usepackage{array,multirow}
\usepackage{caption}

\usepackage{etoolbox}
\makeatletter
\preto{\@verbatim}{\topsep=0pt \partopsep=0pt }

\setlist[itemize]{leftmargin=*,label=\scalebox{.8}{\textbullet}}
\makeatletter
\newcommand{\thickhline}{%
    \noalign {\ifnum 0=`}\fi \hrule height 1pt
    \futurelet \reserved@a \@xhline
}
\newcolumntype{"}{@{\hskip\tabcolsep\vrule width 1pt\hskip\tabcolsep}}
\makeatother

\usepackage[font=small,skip=0pt]{caption}
\usepackage{subfigure}

\makeatletter
\def\@copyrightpermission{\relax}
\def\@mkbibcitation{\relax}
\makeatother

\makeatletter
\let\origsection\section
\renewcommand\section{\@ifstar{\starsection}{\nostarsection}}
\newcommand\nostarsection[1]
{\sectionprelude\origsection{#1}\sectionpostlude}
\newcommand\starsection[1]
{\sectionprelude\origsection*{#1}\sectionpostlude}
\newcommand\sectionprelude{%
  \vspace{.25em}
}
\newcommand\sectionpostlude{%
  \vspace{.25em}
}
\makeatother

\renewcommand\footnotetextcopyrightpermission[1]{}
\begin{document}
\title{Adaptive Processing of Spatial-Keyword Data Over a Distributed Streaming Cluster}

\author{Ahmed R. Mahmood$^1$, Anas Daghistani$^1$, Ahmed M. Aly$^1$, Walid G. Aref$^1$, \\Mingjie Tang$^1$, Saleh Basalamah$^2$, Sunil Prabhakar$^1$}
\affiliation{%
  \institution{$^1$Purdue University, West Lafayette, IN   $\quad \quad \quad ^2$Umm Al-Qura University, Makkah, KSA  }
  \streetaddress{$^1$\{amahmoo,~aaly,~tang49,~aref,~sunil\}@cs.purdue.edu, anas@purdue.edu $\quad \quad \quad ^2$smbasalamah@uqu.edu.sa}
}

\renewcommand{\shortauthors}{Mahmood et al.}
\renewcommand{\shorttitle}{}

\begin{abstract}
The widespread use of GPS-enabled smartphones along with the popularity of micro-blogging and social networking applications, e.g., Twitter and Facebook, has resulted in the generation of huge streams of geo-tagged textual data. Many applications require real-time processing of these streams. For example, location-based e-coupon and ad-targeting systems enable advertisers to register millions of ads to millions of users.
The number of users is typically very high and they are continuously moving, and the ads change frequently as well. Hence sending the right ad to the matching users is very challenging. 
Existing streaming systems are either centralized or are not spatial-keyword aware, and cannot efficiently support the processing of rapidly arriving spatial-keyword data streams. 
This paper presents Tornado, a distributed spatial-keyword stream processing system. Tornado features \textit{routing units} to fairly distribute the workload, and furthermore, co-locate the data objects and the corresponding queries at the same processing units. The routing units use the \textit{Augmented-Grid}, a novel structure that is equipped with an efficient search algorithm for distributing the data objects and queries. Tornado uses \textit{evaluators} to process the data objects against the queries. The routing units minimize the redundant communication by not sending data updates for processing when these updates do not match any query.
By applying dynamically evaluated cost formulae that continuously represent the processing overhead at each evaluator, Tornado is adaptive to changes in the workload. 
Extensive experimental evaluation using spatio-textual range queries over real Twitter data indicates that Tornado outperforms the non-spatio-textually aware approaches by up to two orders of magnitude in terms of the overall system throughput.
\end{abstract}

%
%




\maketitle

\section{Introduction}\label{sec:intro}
Recently, there has been an unprecedented widespread of GPS-enabled smartphones and an increased popularity of micro-blogging and social networking applications, e.g., Twitter, Flickr, and Facebook. In addition, the increased amount of time individuals spend online motivates advertising agencies to keep online traces of the internet users. These online traces include both spatial and textual properties. For example, the online trace of a web search includes both the geo-location and the keywords of each search query. This results in the generation of large amounts of rapidly-arriving geo-tagged textual streams, i.e., spatial-keyword data streams. For example, about 4.4 million geo-tagged tweets and 5 billion Google search queries are generated every day~\cite{wang2015ap,internetstats}. These rapid spatial-keyword streams call for efficient and distributed data processing platforms. 
{
\sloppy
Several applications require continuous processing of spatial-keyword data streams (in real-time). One example is location-aware publish-subscribe systems \cite{wang2015ap}, e.g., e-coupon systems. In these systems, millions of users can subscribe for specific promotions, i.e., continuous queries. For example, a user may subscribe for promotions regarding \textit{nearby} \textit{restaurants and cafes}. Every subscription has a specific spatial range and an associated set of keywords. Each promotion has a spatial location and a textual profile that describes it. An e-coupon is qualified for a user when it is located inside the spatial range of the user's subscription and when the keywords in its textual profile overlap the keywords of the user's subscription. In this application, the number of users and e-coupons can be very high. Users and advertising campaigns can continuously change their target regions and keywords.
}
Another example is real-time event detection and analytics of spatial-keyword data. Users of micro-blogging applications can be viewed as social sensors, where users talk about the events that are happening now. This can help in the real-time detection of events, e.g., accidents, traffic jams, fires, parties, games, etc. 

Despite being in the era of big data, existing systems fall short when processing rapid spatial-keyword streams. These systems belong to one of three categories: (1)~\textit{centralized} spatio-textual systems, e.g.,~\cite{wang2015ap}, that cannot scale to high arrival rates of data, (2)~\textit{distributed} batch-based spatial/spatio-textual systems, e.g.,~\cite{zhang2014efficient,eldawy2015spatialhadoop,aly2015aqwa}, that have high query-latency (where in some cases, it may require several minutes or even hours to execute a single query), and (3)~non-spatio-textual streaming systems, e.g.,~\cite{Sparkstreaming,toshniwal2014storm}, that do not have direct support for spatial-keyword queries. 
This calls for distributed spatial-keyword streaming systems that are equipped with efficient spatial-keyword query evaluation algorithms and structures. 

In this paper, we address the limitations of exiting systems and we describe Tornado~\cite{mahmood2015tornado} a distributed and real-time system for the processing of spatio-textual data streams. Tornado extends Storm~\cite{toshniwal2014storm}. Storm is a distributed, fault-tolerant, and general-purpose streaming system.

Tornado addresses the following challenges:\\
(1)~\textbf{Scalability with respect to data and query workload}: Tornado scales to process a large number of data objects per second against a large number of spatio-textual queries with minimal latency.\\
(2)~\textbf{Skew and variability in workload distribution across time}: It is highly unlikely to have a uniform or a fixed distribution of the data or the query workload. Tornado achieves load balancing, and adapts according to changes in the workload (with minimal overhead).\\
(3)~\textbf{No downtime}: As Tornado adapts to changes in the workload, it is essential to ensure that Tornado is still functional during the transitioning phase, and that the query results are correct, i.e., no missing or duplicate results .\\
(4)~\textbf{Limited network bandwidth}: The underlying network of the computing cluster can easily become a bottleneck under high arrival rates of the data and queries. Tornado minimizes network usage to improve the overall system performance.

To address these challenges, Tornado introduces two main processing layers, namely: 1)~the \textit{ evaluation layer}, and 2)~the \textit{routing layer}.

\noindent
{\bf The Evaluation Layer} is composed of multiple evaluators, where each evaluator is assigned a spatial region, i.e., a \textit{Partition} of the space. The entire space is collectively covered by all the partitions with each partition covering a non-overlapping rectangle.
The routing layer assigns the data objects and the queries to the 
corresponding spatial evaluator(s).

\noindent
{\bf The Routing Layer} distributes data and queries across the processing units, i.e., evaluators. The distribution is location-based, where each evaluator is assigned a spatial region, i.e., a \textit{partition} of the space. One can argue that the distribution of the data and queries can alternatively be text-based. However, text-based distribution is inefficient when compared to location-based distribution. The reason is that a data object, e.g., tweet, has multiple keywords, but only one point location. Text-based distribution may forward a data object to multiple processing units (one per keyword), while space-based distribution forwards a data object to one and only one evaluator.

Employing traditional spatial indexes to achieve location-based distribution of the data is not efficient. For example, a grid index may not be efficient in case of large spatial ranges, while hierarchical spatial indexes, e.g, the quad-tree~\cite{finkel1974quad,samet1990design} or the R-tree~\cite{guttman1984r,beckmann1990r}), require logarithmic time in terms of the total number of processing units. The number of processing units can be large for a large cluster. Furthermore, using location-only distribution does not leverage the textual properties of the data and queries.

The routing layer employs the \textit{Augmented-Grid} (A-Grid, for short), a novel spatial-keyword grid structure. The A-Grid adopts a new algorithm that uses shortcuts to assign data and queries to evaluators. We analytically show that using the A-Grid, the routing time of a query, say $q$, is $O(N_p)$, where $N_p$ is the number of processing units that are relevant to $q$. To reduce the network communication overhead, the A-Grid maintains a textual summary of all the query keywords for every evaluator. Before transmitting a data object, say $O$, to an evaluator, say $A$, the textual summary of $A$ is checked. If the keywords of $A$ do not overlap the keywords of $O$, i.e., $O$ does not contribute to the answer of any query, then $O$ is not transmitted.

\noindent
{\bf Adaptivity}. 
In Tornado, overloaded evaluators can delay the processing and reduce the overall system throughput. Underutilized evaluators waste processing resources. Hence, Tornado maintains a balanced distribution of the workload across all the evaluators. It is expected that the system workload will not be the same at all times, and hence having a static routing layer can result in poor system performance.
Existing systems, e.g.,\cite{aly2015aqwa,kangroo}, address the problem of adaptive workload-aware processing of big data by providing mechanisms for updating the partitioning the data. These systems keep centralized workload statistics, and halt the processing of the data and queries during the re-partitioning phase.
However, in distributed real-time applications, workload statistics are distributed across evaluators and it is unacceptable to pause the query processing. 
This calls for a real-time load-balancing technique that does not interrupt the query processing. It is challenging to implement such a distributed and real-time load-balancing mechanism in Tornado for the following reasons:
\begin{itemize}[noitemsep,nolistsep,leftmargin=*]
\item{\textbf{No Global System View:}} 
In Tornado, the workload statistics are distributed across evaluators. Sending detailed workload statistics from one process to another requires high network overhead. The load-balancing protocol should minimize the overhead needed to collect, transfer, and process the workload statistics. 
\item{\textbf{Correctness of Evaluation:}} during the re-partitioning phase, Tornado redefines the boundaries of the evaluators. This requires moving queries from one evaluator to another.
Meanwhile, the data objects continuously update their locations, and the answer to each query needs to be continuously updated as well. Hence, unless the incoming data objects are carefully directed, missing (or duplicate) results can occur.
\item{\textbf{Overhead of Re-partitioning:}} Moving the queries between the evaluators incurs network overhead. The re-balancing algorithm should be aware of the re-balancing overhead, and avoid unnecessary re-balancing.
\end{itemize}

Tornado employs a \textit{decentralized} load-balancing mechanism, where the choice of the new spatial boundaries of the evaluators is delegated to the evaluators themselves. This reduces communication overhead needed to transfer detailed workload statistics and distributes the computational overhead across the evaluators.
The load-balancing mechanism is incremental, i.e., rather than redefining all the partitions, only a few partitions are updated using simple shift, split, and merge operations. Furthermore, Tornado ensures the correctness of evaluation during the transient phase using a two-stage re-partitioning protocol.


\noindent
In summary, the contributions of this paper are as follows:
\begin{itemize}[noitemsep,nolistsep]
\item We introduce Tornado, a scalable spatio-textual data streaming system.
\item We develop an \textit{Augmented-Grid} structure and an optimal \textit{neighbor-based routing} algorithm that minimizes the overhead of routing the data and queries. The routing layer is \textit{spatio-textual} and prohibits routing data objects with no matching queries and minimizes network overhead.

\item We present an  \textit{incremental}, \textit{adaptive}, and  \textit{decentralized} load-balancing mechanism that ensures fairness in the workload distribution across the evaluators.
\item Using real datasets from Twitter, we show that Tornado achieves performance gains of up to two orders of magnitude in comparison to a baseline approach.
\end{itemize}
The rest of this paper proceeds as follows. Section~\ref{sec:preliminaries} presents the notations used throughout the paper.
Section~\ref{sec:tornadostructure} describes the structure of Tornado.
Section~\ref{sec:adaptivity} describes the load balancing mechanism in Tornado.
Section~\ref{sec:analysis} formally analyzes how to set Tornado's system parameters.
The related work is presented in Section~\ref{sec:relatedwork}.
Detailed experimental evaluation is given in Section~\ref{sec:experimentalevaluation}.
Section~\ref{sec:conculsion} contains concluding remarks.

\section{Preliminaries}\label{sec:preliminaries}
In this section, we present the notations that are used throughout the paper. 
A spatial-keyword data stream is an unbounded sequence of spatial-keyword objects. A spatial-keyword object, say $O$, 
has the following format:
$O=\left[oid,~loc,~text,~ts\right]$, where $oid$ is the object identifier, $loc$ is the geo-location of the object at Timestamp $ts$, and $text$ is the set of keywords associated with the object.

  

\textbf{A continuous spatial-keyword filter query}, say $q$, is defined as $q=\left[qid,~MBR,~text,~t\right]$, where $qid$ is the query identifier, $MBR$ is minimum bounding rectangle representing the spatial range of the query, and $text$ is the set of keywords of the query.
The continuous query $q$ is registered, i.e., keeps running for a specific duration, say $t$. During $t$, the query continuously reports the data objects that satisfy the query's spatial and textual predicates. To satisfy a query, a data object needs to be located inside the spatial range of the query, and needs to satisfy the textual predicate of the query. 
{\sloppy
In general,
Tornado supports the following spatial-keyword constructs:
\begin{itemize}[noitemsep,nolistsep]
\item \textit{INSIDE(MBR)}: This spatial predicate evaluates to True when the location of the object is inside $MBR$, i.e., the area that represents the minimum bounding rectangle of the spatial range of the query.
\item \textit{OVERLAPS(text1,~text2)}: This predicate evaluates to True when there is an overlap between the keywords of \textit{text1} and the keywords of \textit{text2}, e.g., \textit{text1}=\{``food",~``sale",~``coupon"\} and \textit{text2}=\{``cafe",~``food",~``restaurant"\}, because the keyword ``food" is shared between \textit{text1} and \textit{text2}.
\item \textit{CONTAINS(text1,~text2)}: This predicate evaluates to True when all the keywords of \textit{text2} are contained in the keywords of \textit{text1}, e.g., \textit{text1}=\{``food",~``sale",``coupon"\} and \textit{text2}=\{``sale",``food"\}, because the keywords ``sale" and ``food" of \textit{text2} exist in \textit{text1}.
\end{itemize}}
Figure~\ref{fig:publishsubscribe} gives an example of multiple spatial-keyword filter queries from a publish-subscribe e-coupon application. We use this example throughout the rest of the paper. An e-coupon is qualified for user's subscription if it is located inside the spatial range of the subscription and when the textual profile of the e-coupon matches with the textual predicate of the subscription. In Figure~\ref{fig:publishsubscribe}, three subscriptions, i.e., queries,  $q_1,q_2,$ and $q_3$ are registered in the system. E-coupon $o_1$ qualifies for subscription  $q_3$ because it is located inside the spatial range of $q_3$  and  the textual content of $o_1$, i.e., ``free, coffee, sandwich" contains the keywords of $q_3$, i.e., ``free, coffee". 
\section{Tornado System Architecture}\label{sec:tornadostructure}
In this section, we present the architecture of Tornado, and its main processing units along side with query processing algorithms.
Tornado~\cite{mahmood2015tornado} extends Storm~\cite{toshniwal2014storm}. 
Storm is a cluster-based, distributed, fault-tolerant, and general-purpose streaming system that achieves real-time processing with high throughput and low latency. Storm provides three abstractions, namely: \textit{spout}, \textit{bolt}, and \textit{topology}. A spout is a source of input data streams. A bolt is a data processing unit. A topology is a directed graph of bolts and spouts that resembles a pipeline of streamed data evaluation.

Storm is not optimized for the execution of spatial-keyword queries, simply because it does not have built-in support for spatial or textual primitives, e.g., points, rectangles, or overlap/containment of keyword lists.

In order to efficiently support the evaluation of spatial-keyword queries, we need to guarantee that relevant data and queries are collocated in the same processing unit, i.e., a Storm bolt. 
This is challenging because the system needs to distribute data and queries across processing units in a way that achieves the following properties: (1)~{\bf Optimize the  network communication overhead} within the cluster by not sending the same data object to multiple processing units, (2)~{\bf Optimize the memory usage} across the machines by not storing queries in multiple processing units, (3)~{\bf Optimize the CPU usage} by checking each data object against as few queries as possible, and (4)~{\bf Maintain good load balancing as the workload changes}, and distribute the data and queries across the processing units while guaranteeing the correctness of evaluation, i.e., without missing output tuples and without producing duplicate results.
\begin{figure}[t]
\centering
\includegraphics[width=3.3in]{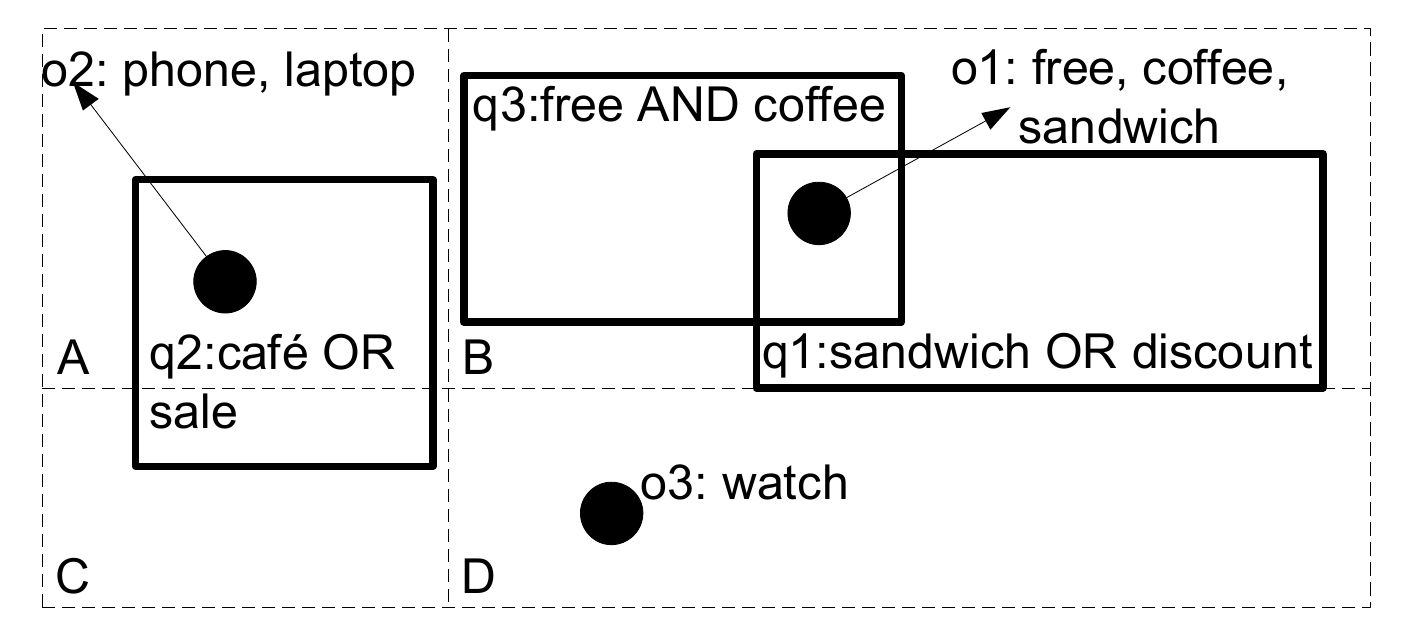}
\caption{Assigning coupons to customers according to spatio-textual overlap.}
\label{fig:publishsubscribe}
\end{figure}
\begin{figure*}[!t]
\captionsetup[subfigure]{justification=}
        \subfigure[The routing units and evaluators.]{	\includegraphics[width=2.1in]	{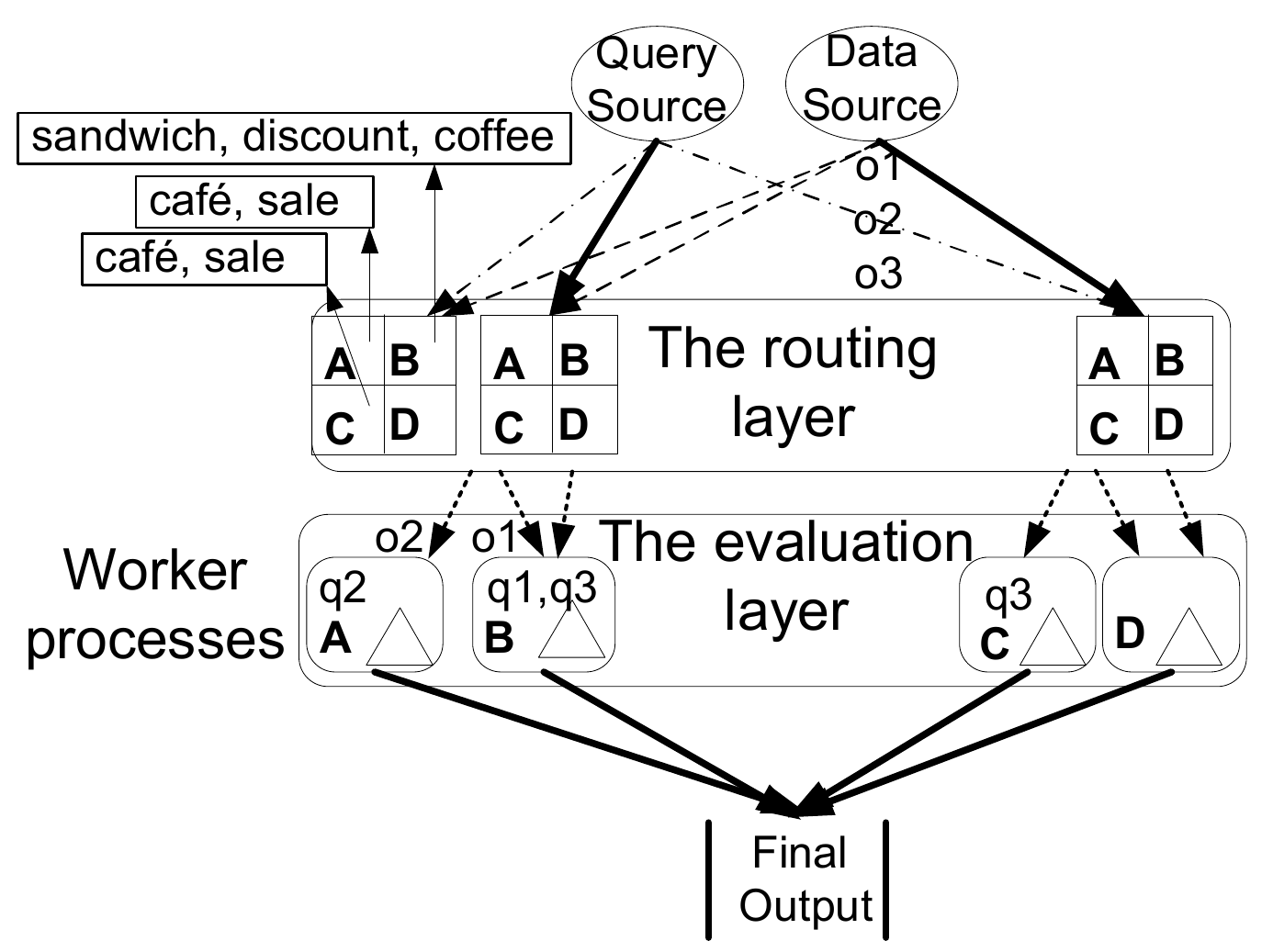}}  
          \subfigure[Spatio-textual indexing.]{	\includegraphics[width=1.7in]	{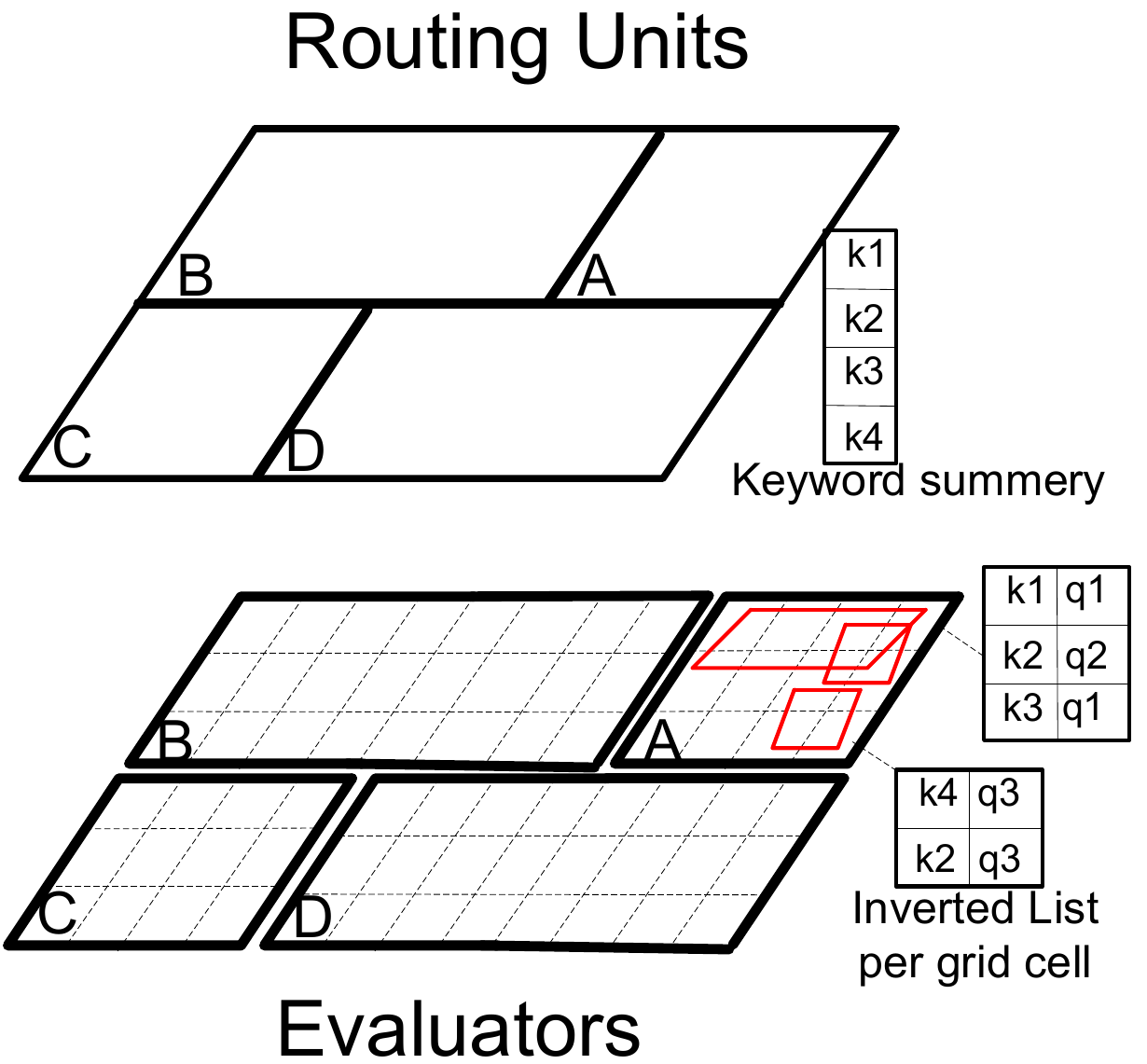}}
             \subfigure[Neighbor-based query routing in Tornado.]{	\includegraphics[width=2.5in]	{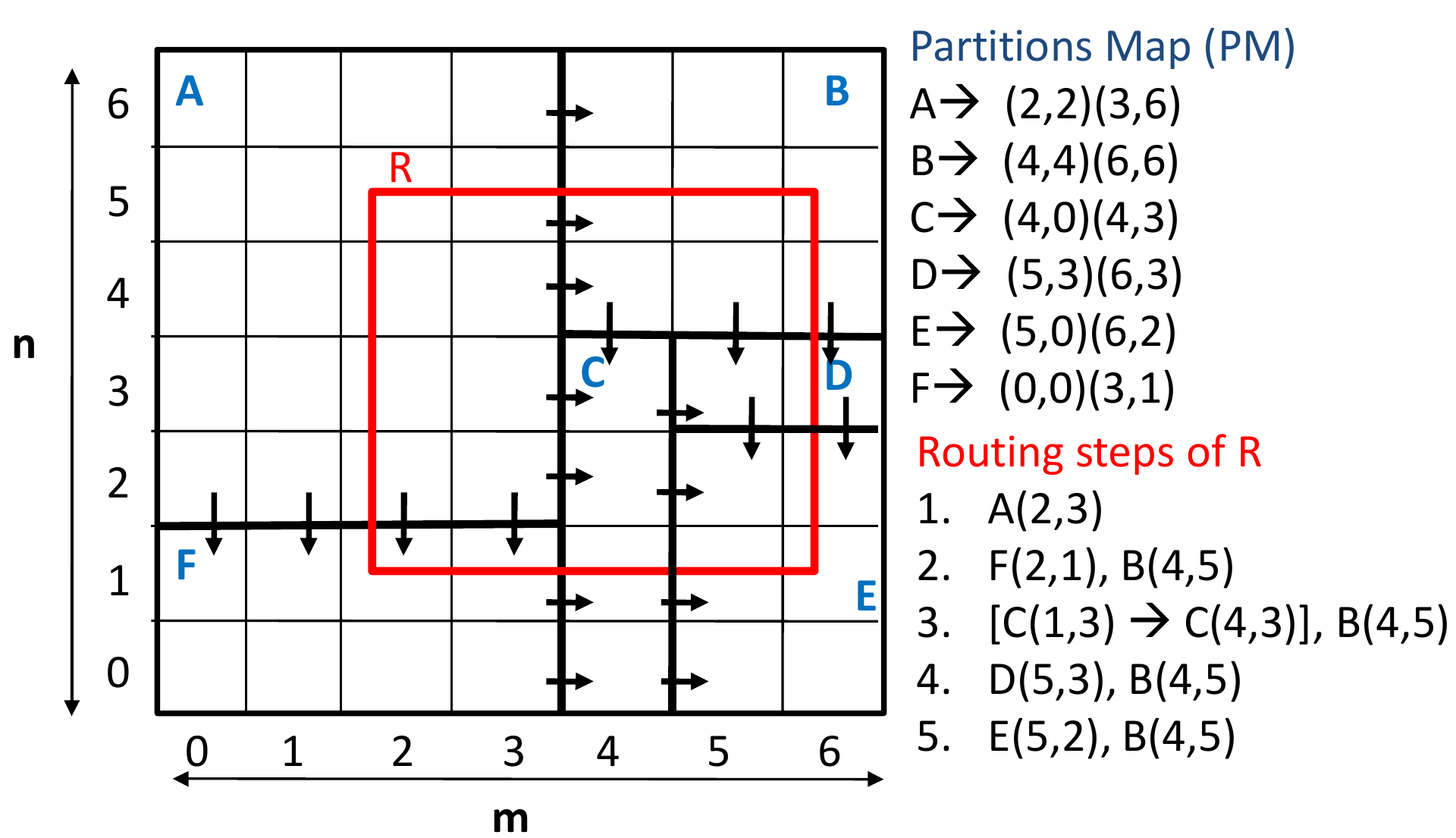}}       
        \caption{The architecture and system components of Tornado. }\label{fig:stormvstornado}
\end{figure*}
 
Tornado addresses the above challenges by co-locating the data objects with the relevant queries.
Tornado extends the bolt abstraction from Storm into \textit{routing units} and \textit{evaluators}. The routing units are light-weight components that are responsible for co-locating the queries and data objects together. The evaluators are processing units that check the incoming data objects against the continuous queries and produce query results.

Tornado makes use of the fact that a data object has a single point location, but multiple keywords. This is typical in many location services, e.g., as in tweets, where a tweet is associated with a single location and multiple keywords. Accordingly, the routing layer in Tornado partitions the space into non-overlapping MBRs. Every evaluator is responsible for a single MBR. The benefit of having non-overlapping MBRs is to optimize the network utilization by forwarding each data object to a single evaluator. 

To support high arrival rates of streamed data, the routing layer applies replication, i.e., multiple identical routing units are employed. 
The routing 
layer
maintains a textual summary for every evaluator. The textual summary of an evaluator, say $E$, contains all keywords of queries stored in $E$.  
In the routing units, the textual summary for an evaluator is stored as a \textbf{hash table} of keywords.
Before forwarding a data object, say $O$, to an evaluator, say $E$, the textual summary of $E$ is consulted to check if there are some queries in $E$ that have keywords that overlap the keywords of $O$. 
Figure~\ref{fig:stormvstornado}(a) illustrates how Tornado processes the queries of Figure~\ref{fig:publishsubscribe}. 
Once a query is received, a routing unit is selected at random, 
and the query is forwarded to that routing unit, where the latter sends the query to the spatially relevant evaluator(s). Based on the textual summary of the evaluators, stored on the routing layer,
some data objects are not forwarded to any evaluator, e.g., $o_3$ in Figure~\ref{fig:stormvstornado}(a).

\subsection{The Routing Units: The Augmented-Grid (A-Grid)}
The routing layer is composed of multiple identical routing units. An instance of the routing units maintains a spatial-keyword index to properly route the data objects and the queries. 
In terms of spatial indexing, 
a routing unit partitions the entire space into a virtual fine grid $FG$. 
Then, the space is partitioned into $N_e$ non-overlapping spatial partitions that are overlaid on top of the fine grid.
Each partition, say $p$, corresponds to one evaluator, and is defined as follows: $[pid, xcellmin, ycellmin, xcellmax,ycellmax]$, where $pid$ is the identifier of the partition, $xcellmin$ and $ycellmin$ define bottom left grid cell of $p$, $xcellmax$ and $ycellmax$ define the top right grid cell of $p$. Every routing unit maintains a summary of the query keywords per evaluator.

Tornado employs the fine grid partitioning for two reasons: (1)~To speed-up the routing time, and (2)~To support the rearrangement of the boundaries of the evaluators during the load-balancing procedure (that is explained in Section~\ref{sec:adaptivity}).

An incoming data object or query goes to a random instance of the routing units to be assigned to the corresponding evaluator(s). The smaller the routing time, the higher the throughput of the entire system. Moreover, having light-weight routing units can save more resources that can be used for query evaluation rather than for routing. In Tornado, the location of a data object is represented as a single point in space. Because the partitions are non-overlapping, a data object is routed to a single evaluator. This routing is achieved in $O(1)$ using uniform grid partitioning. However, a query has a spatial range, that can overlap multiple partitions, and hence a query needs to be routed to multiple evaluators.

To find the evaluators, a data object or a query belongs to, one can index the partitions using a traditional structure, e.g., a grid or an R-tree. However, these structures are not efficient when adopted in the routing layer of Tornado. For instance, using a spatial grid to index the spatial partitions of the evaluators is not efficient for queries with large spatial ranges. The reason is that in order to identify all the partitions to which a spatial range belongs, we need to traverse all the grid cells that overlap the spatial range of the query. This may require visiting many redundant grid-cells that belong to the same partition. This process takes $O(n\times m)$ time, where $n\times m$ is the total number of grid cells to be touched in the worst case. The finer the granularity of the grid, the higher the search overhead. Furthermore, using a hierarchical index, e.g., an R-tree, requires $O(\log N_e+N_p)$~\cite{guttman1984r,de2000computational} routing time, where $N_e$ is the overall number of evaluators in the system, and $N_p$ is the number of evaluators that match an incoming data object or a query.
\begin{mydef}\textbf{The Routing Problem}
Given a rectangular query-range, say $r$, and a set, say $S$, of $N_e$ non-overlapping rectangular partitions that cover the entire space, find the partitions that overlap $r$.
\end{mydef}
We propose \textit{Neighbor-Based Routing}, a novel and optimal routing technique that requires 
$O(N_p)$ operations to route a spatial range,
where $N_p$ is the number of evaluators that overlap the spatial range. This is lower than the time needed in both the traditional grid, i.e., $O(m\times n)$ and hierarchical structures, i.e., $O(\log N_e+N_p)$.

The main idea of neighbor-based search algorithm is to follow shortcuts to jump directly from \textit{dominant cells} belonging \textit{neighboring} partition. 
\begin{mydef}\textbf{Dominant cell}
A dominant cell of a partition, say $A$, with respect to a spatial range, say $R$, is the top left cell of $A$ that is inside $R$.
\end{mydef}
For example in Figure~\ref{fig:stormvstornado}(c), the dominant cell of the Partition $A$ with respect the spatial range $R$ is (2,5). 
Observe that each grid cell is spatially contained inside the spatial range of a single evaluator. Boundaries of partitions are maintained an a hash table termed the \textit{Partitions Map}, PM for short as illustrated in  Figure~\ref{fig:stormvstornado}(c). 

Each grid cell, say $c$, maintains the identifier of the partition that contains $c$. To find the right dominant cell with respect to a range $R$ we follow the following steps:(1)~find the right cell $RC$ belonging to a different partition, and (2)~find the dominant cell of $RC$ that is the top-left of $RC$ belonging to the same partition and inside the spatial range $R$.
Refer to Figure~\ref{fig:stormvstornado}(c) for illustration. Assume that we need to identify the right dominant cell, say $RC$ of Cell $(2,5)$ within Partition $A$. From the PM we know that the partition $A$ spans cells [$(0,5),(3,6)$]. The right cell $RC$ of the cell $(2,5)$ is
of the form $(xp,yp)$, where $xp$ is the index on the horizontal coordinate that is to the right of cell  $(2,5)$. The $yp$ is $5$ because the Cell $RC$ is to the right of Partition $A$ and has the same position on the vertical axis. From the PM, the partition $A$ ranges from $2$ to $3$ on the horizontal coordinate, where $2$ and $3$ are $xmin$ and $xmax$ of the Partition $A$ respectively, the value $xp$ is equal to $4$ that is $1+xmax$. This means that the Cell $RC$ is $(4,5)$ that is covered by Partition $B$. The dominant cell of $(4,5)$ is also $(4,5)$ as this is the top-left cell within $R$. The same logic applies when finding the bottom dominant cell.

To route a spatial range, say $R$, we start from the upper-left corner of $R$. We find the partition that is covered by that corner (this is trivial because the partition identifier is stored in the cell corresponding to that corner). Then, we follow the right and bottom dominant shortcuts of that corner.
We recursively apply this procedure until we reach a cell from which the pointers lead to a cell that is outside $R$ or to a previously visited partition. We use a Boolean array to mark the visited partitions and avoid visiting the same partition more than once.
Refer to Figure~\ref{fig:stormvstornado}(c) for illustration. To route the red rectangle, we start from Cell $(2,5)$ covered by Partitions $A$. Then, we follow the pointers to Cells $(4,5)$ and $(2,1)$, covering Partitions $F$ and $B$, respectively. Then, we follow the bottom pointer of Cell $(2,1)$ to reach Cell $(4,1)$ inside Partition $C$. From the PM, we identify that Cell $(4,3)$ is the dominant cell of the Partition $C$ with respect to $R$. We follow dominant cell shortcuts visiting the following cells: Cell $(5,3)$ within Partition $D$, Cell $(5,2)$ within Partition $E$, and Cell $(4,5)$ within Partition $B$. The Pseudocode of the algorithm is given in Algorithm~\ref{alg:findEValuators}. 
\begin{algorithm}[h]
Stack\ S\\
Cell\ c(x,y)$\gets$ TopLeft corner of r\\
S.push(c)\\
\While{S\ \textbf{not}\ empty}{
	c$\gets$S.pop\\
    \If{c\ overlaps\ r\ \textbf{and}\ c.partition is not visited}
    {
     add c.partition to result\\
     mark c.partition as visited\\
     rightCell = getDom(getRightCell(c.y))\\
     bottomCell = getDom(getBottomCell(c.x))\\
      S.push(bottomCell),S.push(rightCell)\\
     }
}
\caption{$neighborSearch$(MBR r)}\label{alg:findEValuators}
\end{algorithm}

One can think of the A-Grid as a hypothetical directed-acyclic-graph (DAG), where nodes of the graph are A-Grid cells and edges are the right and bottom shortcuts to neighbour dominant cells belonging to different partitions. The neighbor-based search can be seen as performing a special type of traversal on the cells of the A-Grid where only the dominant cells inside the partitions overlapping the spatial range of the query are visited. \\


\begin{mylemma}\textbf{The neighbor-based routing requires $O(N_p)$ and does not depend on the granularity of the grid}\\
For the traversal performed by the neighbor-based search algorithm, the number of nodes $V$ in the hypothetical DAG is $N_P$. The number of the edges $E$ visited is $2N_P$ because for every node, we follow at most two pointers. The total traversal time is $O(V+E)=O(N_p)$. The run time of the algorithm cannot be less than $O(N_p)$ as this is the size of the output.
\end{mylemma}

The initialization phase requires $O(n\times m)$ to assign partition identifiers to all cells with the A-Grid, where $n\times m$ is the total number of A-Grid cells, and $n$ and $m$ are the number of cells in the $x$ and $y$ axes, respectively.



For an incoming data object, say $O$, after the relevant evaluator, say $E$, is determined, Tornado considers the textual contents of $E$. If none of the queries that are registered at $E$ share any keywords with $O$, then $O$ is not routed to $E$. 
To achieve this, Tornado maintains in the routing units, a summary of query keywords within each evaluator. 
As described in Section~\ref{sec:preliminaries}, given a query say $q$, Tornado supports two types of textual predicates, namely \textit{OVERLAPS} and  \textit{CONTAINS}.
On the one hand, a data object satisfies an \textit{OVERLAPS} predicate if that object has any of the keywords of $q$. This requires that all the keywords of $q$ exist in the textual summary of the evaluator corresponding to $q$.
On the other hand, a data object satisfies a \textit{CONTAINS} predicate if that object has all of the keywords of the query $q$. Tornado reduces the size of the textual summary and the communication needed for the \textit{CONTAINS} textual predicate. In this case,
as a filtering step, Tornado stores only a single keyword from $q$ in the textual summary of the evaluator corresponding to $q$. This approach achieves up to 5 times higher query throughput, as illustrated in Section~\ref{sec:experimentalevaluation}.

Observe that Tornado maintains multiple identical routing units. One way to keep track of the query keywords within each evaluator is to \textit{broadcast} each query to all the routing units. To avoid unnecessary communication, an incoming query, say $q$, goes to an arbitrary instance of the routing units, say $U$. If $q$ adds new keywords to any evaluator, say $E$, then $U$ \textit{forwards} the added keywords to the other replicas of the routing units. As queries expire, the textual summary of the evaluators may contain redundant keywords. We describe how to update the textual summary in Section~\ref{subsec:falsepositives}.

\begin{figure*}[t!]
 \begin{minipage}{.7\textwidth}
        \subfigure[Before split/merge.]{	\includegraphics[width=1.5in]			{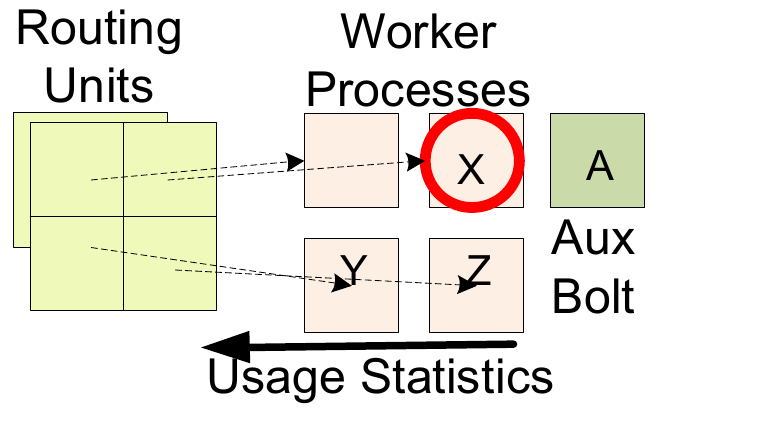}}
        \subfigure[Transient phase.]{	\includegraphics[width=1.5in]				{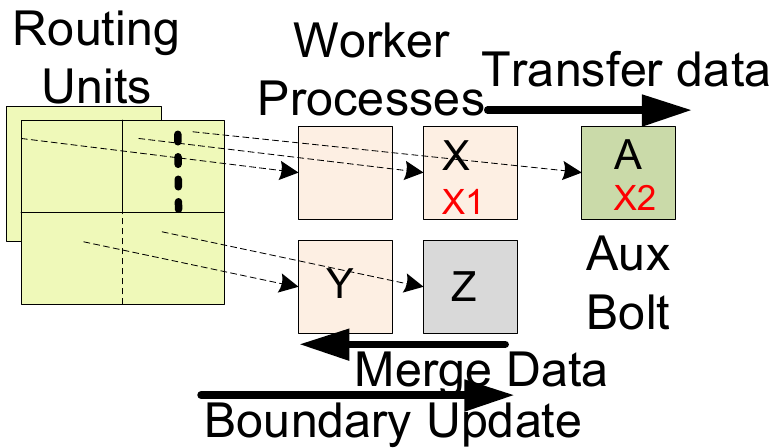}}
        \subfigure[After split/merge.]{	\includegraphics[width=1.5in]				{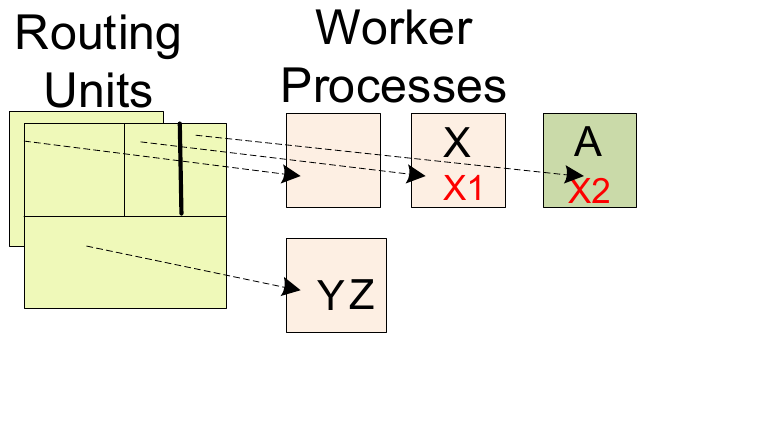}}
        \caption{The split/merge operation.}\label{fig:splitmerge}
        \end{minipage}%
        \begin{minipage}{.3\textwidth}
  \includegraphics[width=1.8in]{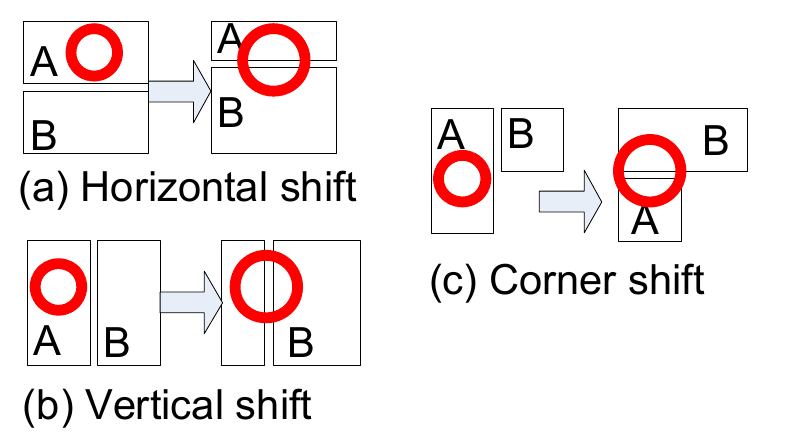}
       \caption{Shift variations.}\label{fig:shiftcases}
   \end{minipage}%
\end{figure*}
\subsection{Evaluators}\label{subsec:evalautor}
To improve the overall system performance, each evaluator maintains a spatio-textual index. In particular, each cell in the fine grid maintains an inverted list with the keywords of the registered continuous queries.
Figure~\ref{fig:stormvstornado}(b) describes the indexes adopted within each evaluator. 
The main responsibilities of an evaluator are as follows:
\begin{enumerate}[noitemsep,nolistsep,leftmargin=*]
\item Store and index continuous queries and drop expired queries.
\item Process incoming data objects against stored queries.
\item Keep track of usage and workload statistics. 
\end{enumerate}

To register a continuous query into an evaluator, first, we find the grid cell(s) that overlap the spatial range of the query. Then, at each overlapping cell, we attach the query to the inverted list.
To process an incoming data object, say $O$, we find the grid cell, say $C$, that contains $O$'s location. Then, using $O$'s keywords and $C$'s inverted list of queries, we retrieve a list of candidate queries that have $O$'s keywords. Finally, we verify if $O$ belongs to the answer of each of these candidate queries. 
The final verification phase is both spatial and textual. For every query, in the candidate query list, we fist check if it contains the data object, then we check the textual predicate of the query. If the textual predicate of the query is $OVERLAPS$, then there is no need to further verify the textual predicate. If the textual predicate is $CONTAINS$, then we must verify that the data object fully contains all keywords in the query.
\section{Real-time Load Balancing}
\label{sec:adaptivity}
In Tornado, each evaluator is responsible for a certain spatial range that covers a partition in the fine grid $FG$. To achieve high throughput, Tornado keeps a balanced distribution of the workload across the evaluators.
To compute the workload corresponding to an evaluator, Tornado keeps workload statistics at the grid cells of $FG$. 
For each data object, say $O_l$, that is received by $FG[i][j]$, where $i$ and $j$ are the horizontal and vertical coordinates of the Cell $FG[i][j]$, respectively, let $q_l$ be the number
of queries that contain any of the keywords of $O_l$. 
Observe that the number of queries that contain a certain keyword can be easily retrieved from the inverted list within each grid cell.
For each grid cell $FG[i][j]$, we define the workload overhead, i.e., the computational cost,  as the sum of $q_l$ over all the data objects $O_l$ received by that cell:
\begin{equation}
    cost(FG[i][j]) =\sum_{l}  q_l
    \label{eq:singlecellcost}
\end{equation}
{\sloppy Given a partition, say $P_w$, that is bounded by $[({xmin},~{ymin}),~({xmax},~{ymax})]$, the overall computational cost is the sum of the costs of all the grid cells in $P$, i.e.,
\begin{equation}\label{eq:partitioncost}
cost(P_w)=\sum cost(FG[i][j])
\end{equation}
where ${xmin}\leq i \leq {xmax}$ and ${ymin}\leq j \leq {ymax}$. 
}
Below, we describe the load-balancing protocol in Tornado.

\subsection{Initialization}\label{subsec:initialzation}
Tornado partitions the entire space into $N_e$ partitions, where $N_e$ is the number of evaluators.
To choose the initial boundaries of the partitions, Tornado uses a sample of the data and query workload, and calculates the computational cost of each fine grid cell. Let $\alpha$ 
be
the maximum computational cost of the partition $P_w$,
i.e., 
\begin{equation}\label{eq:maximization}
\alpha=\max_{P_w}(cost(P_w))
\end{equation}
In the initialization phase, the objective is to minimize $\alpha$ across all the $N_e$ partitions. The best-case distribution is to have all evaluators 
process
equal portions of 
the  
workload. 
The problem of finding the optimal rectangular partitioning that minimizes $\alpha$ is NP-Hard (see~\cite{grigni1996complexity}).
Tornado employs a hierarchical recursive space decomposition similar to that of a k-d tree decomposition~\cite{ooi1987spatial,aly2015aqwa}.
In particular, Tornado maintains a priority queue of the partitions to be split, where the partitions are sorted according to their cost. First, the entire space represents a single partition 
that
is inserted 
into 
the priority queue. Then, the top partition from the queue, i.e, the one with the highest cost, is retrieved, and then is split into two partitions. The split is chosen in a way that minimizes the maximum cost of the 
resulting 
two 
sub-partitions. Then, the resulting sub-partitions are inserted into the priority queue. This process is repeated until a single grid cell is reached (that cannot be split), or the maximum allowed number of 
evaluators
in the system is reached. The maximum number of evaluators is a system parameter that affects the performance of Tornado.  As 
illustarted 
in Section~\ref{sec:experimentalevaluation}, having few evaluators results in an under-utilized cluster. Also, having more evaluators than the cluster resources results in contention among the evaluators, which degrades the performance. 

\subsection{Adaptivity in Tornado}
To preserve fairness in workload distribution while keeping the number of evaluators fixed, Tornado uses two \textit{incremental} load-balancing operations, namely: (1)~\textit{split/merge}, and (2)~\textit{shift}. 

\noindent
A \textbf{split/merge} operation involves a split of an overloaded evaluator into two evaluators, followed by a merge of two neighboring underutilized evaluators into a single evaluator. The split is either horizontal or vertical. The split position is chosen to minimize the difference in cost between the resulting two partitions. The details of finding the best point to split in an evaluator are given in Section~\ref{subsec:decenterlized}. During a split, Tornado transfers some grid cells from an overloaded evaluator to an auxiliary evaluator.
Refer to Figure~\ref{fig:splitmerge} for illustration. Figure~\ref{fig:splitmerge}(a) illustrates an overloaded evaluator $X$ before a split/merge operation. An instance of the routing units makes a decision to split/merge and initiates a split of Evaluator $X$ into $X_1$ and $X_2$, and a merge of Evaluators $Y$ and $Z$, as 
in Figure~\ref{fig:splitmerge}(b).  Observe that, according to the new boundaries, some of the fine grid cells are being transmitted from evaluator $X$ to an auxiliary evaluator $A$. All the fine grid cells that are stored in Evaluator $Z$ are transferred to Evaluator $Y$. Figure~\ref{fig:splitmerge}(c) gives the state at the end of the split/merge operation. 

A \textbf{shift} operation involves a transfer of the workload, i.e., fine grid cells, from an overloaded evaluator to an underutilized spatially adjacent evaluator. The shift operation is useful when no merge of two lightly loaded evaluators is possible.
Tornado uses three variants of the shift operation, namely: \textit{horizontal}, \textit{vertical} and \textit{corner} shifts.
Refer to Figure~\ref{fig:shiftcases} for illustration. The red circle in the figure represents an area with high workload. A horizontal shift is applicable to two evaluators that share a horizontal boundary, e.g., see Figure~\ref{fig:shiftcases}(a). Similarly, a vertical shift is applicable to two evaluators that share a vertical boundary, e.g., see Figure~\ref{fig:shiftcases}(b).
A corner shift is applicable when two neighboring evaluators form a corner shape, e.g., see Figure~\ref{fig:shiftcases}(c). The corner shift allows a transfer of workload between two non-mergeable evaluators, i.e., ones that do not share an entire horizonal or vertical boundary. The details for finding the best point to shift are described in Section~\ref{subsec:decenterlized}.

The neighborhood information among the evaluators is determined during the initialization phase of augmenting the grid as described in Section~\ref{subsec:initialzation}. 
After rebalancing,
the neighborhood information is updated as described in Section~\ref{subsec:updates}.

The decision of whether to initiate a rebalancing operation or not depends on two factors, namely, the \textit{cost reduction \textbf{$C_r$ }} resulting from the re-balance
operation, and the \textit{cell transfer overhead \textbf{$C_t$ }} involved in the re-balance operation.
{\sloppy
\noindent
\textbf{The cost reduction \textbf{ $C_r$ }} of a re-balance operation is the difference between the maximum partition cost before and after the re-balance operation. Consider the split/merge operation 
in Figure~\ref{fig:splitmerge}, and assume that Evaluator $X$ 
has
the highest cost. The cost before split/merge=$cost(X)$. 
The cost after split/merge is $max(cost(X_1),cost(X_2),(cost(Y)+cost(Z)))$. The cost reduction of the split/merge operation is: 
\begin{equation}\label{eq:splitmergereduction}
\begin{split}
&C_r(split/merge, X,X_1,X_2,Y,Z)=\\
&cost(X)-max(cost(X_1),cost(X_2),(cost(Y)+cost(Z)))
\end{split}
\end{equation}
The above idea applies to the shift operation, where the cost reduction is computed as the difference between the maximum cost before and after the shift operation.
 }
\noindent
\textbf{The cell transfer overhead \textbf{ $C_t$} } is an estimate of the overhead of transferring cells during the re-balance operation. $C_t(p)=\beta \times queryCount(p)$, where $queryCount(p)$ is the number of queries in Partition $p$, and $\beta$ is the average time needed to transfer a query. $queryCount(p)$ is incremented whenever a query is registered at $p$, and is decremented whenever a query in $p$ expires.
For example, for the split/merge operation 
in Figure~\ref{fig:splitmerge}, the cell transfer overhead of the split/merge operation is calculated as follows: 
\begin{equation}\label{eq:splitmergetrnasfer}
\begin{split}
&C_t(split/merge, X,X_1,X_2,Y,Z)=\\
&\beta \times(queryCount(X_2)+queryCount(Z))
\end{split}
\end{equation}
Tornado chooses the operation that maximize that value of $C_r$ while having $C_r>C_t$.

\subsection{Decentralized Load-Balancing}\label{subsec:decenterlized}
Existing load-balancing approaches are centralized~\cite{aly2015aqwa,kangroo}, i.e., require having a single unit that receives all the workload statistics. In contrast, in Tornado, the computation of the costs of the fine grid cells is distributed across the evaluators.
The evaluators keep detailed workload statistics and choose the split coordinates that are needed to perform the shift and split/merge operations.
The routing layer periodically receives a summary of the workload statistics from the evaluators, and then makes a decision 
as to whether 
to change the partitioning or not. Also, the routing layer decides which operation to perform. 

For the routing layer to make a decision whether to re-balance or not, it does not need the detailed costs of every grid cell. 
The decision to rebalance can be made using the overall evaluator costs from Equation~\ref{eq:partitioncost}. 
The rebalancing approach adopted in Tornado is decentralized for the following reasons: 1)~Accurate workload statistics are distributed across evaluators, 2)~Each evaluator independently chooses the optimal set of grid cells to be transferred to improve throughput, and 3)~The routing layer only makes a decision whether to change the partitioning or not.

\begin{figure}
 		\centering
        \subfigure[Initial evaluator statistics.]{	\includegraphics[width=1.6in]	{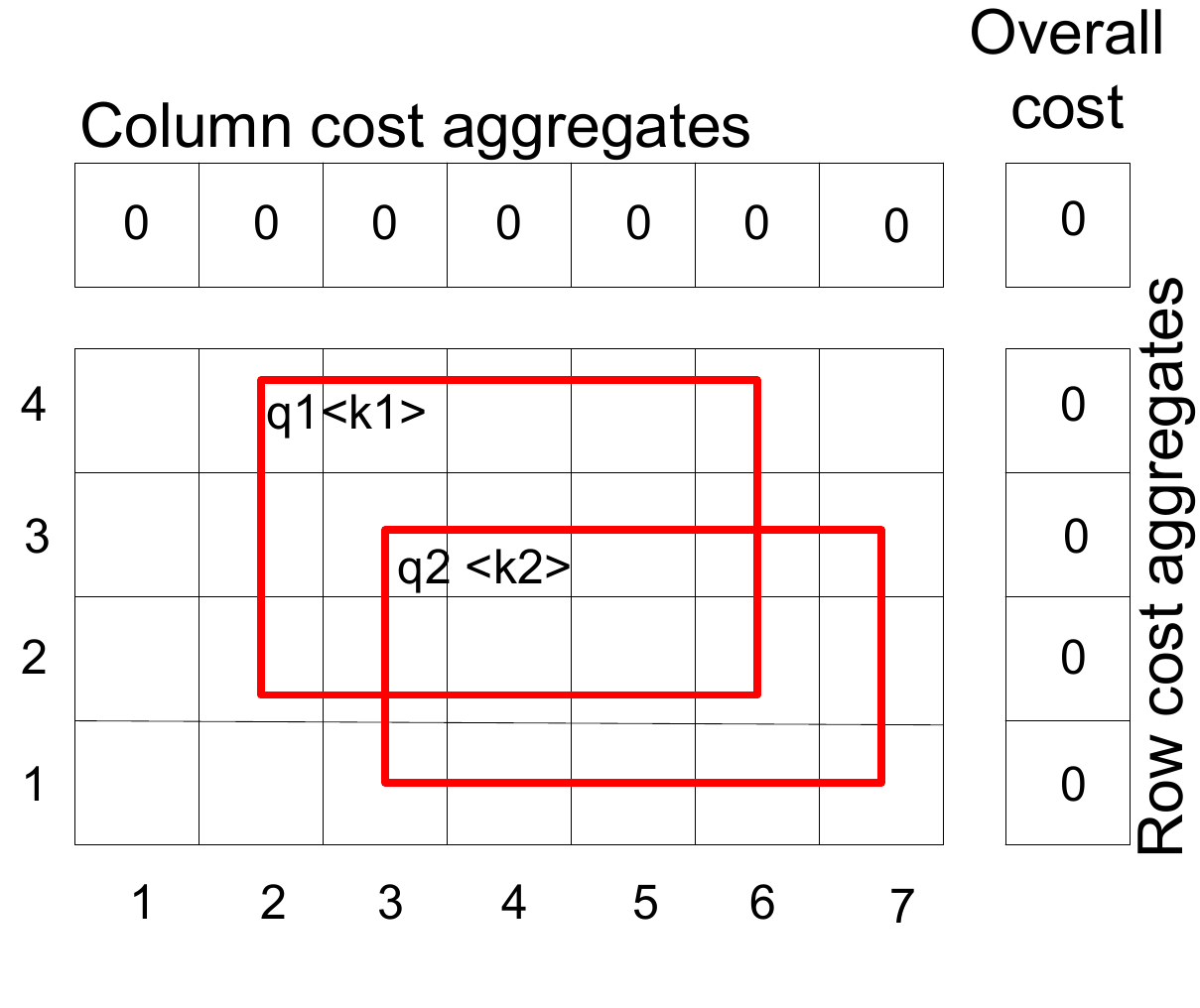}}
        \subfigure[After data objects arrive.]{	\includegraphics[width=1.6in]	{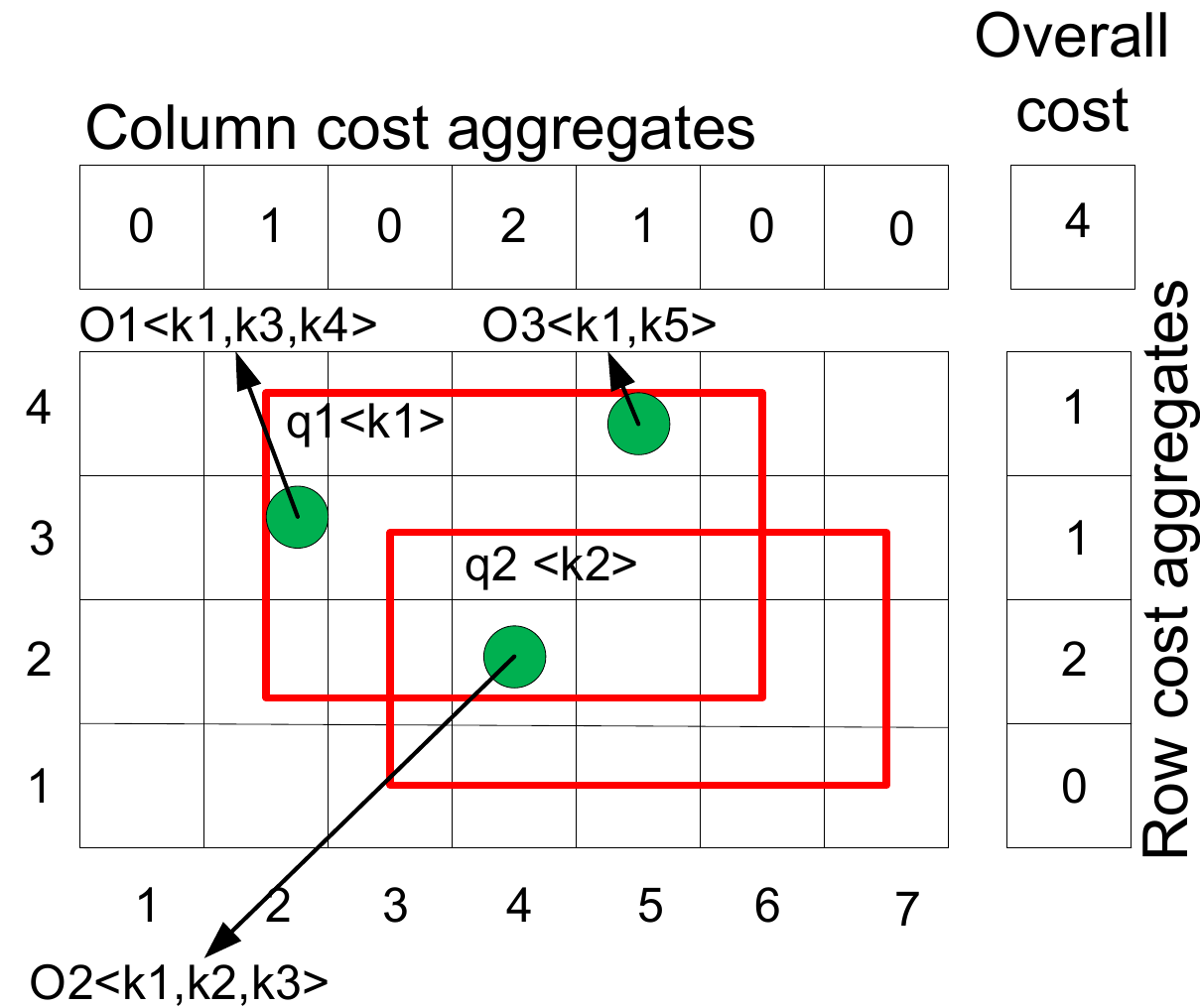}}  
	  \caption{Cost aggregation within an evaluator.   }\label{fig:finegridcost}
\end{figure}


Tornado keeps three aggregates at each evaluator, namely, row, column, and overall aggregates.
Refer to Figure~\ref{fig:finegridcost} for illustration.
Figure~\ref{fig:finegridcost}(a) 
gives 
the initial values of these aggregates.
Figure~\ref{fig:finegridcost}(b) 
gives 
the values after processing three data objects $O_1$, $O_2$, and $O_3$. $O_1$ satisfies one query at Cell $(2, 3)$, and hence the aggregates of Row~3 and Column~2 are incremented. $O_2$ satisfies two queries at Cell $(4,2)$, and hence the aggregates of Row~2 and Column~4 increase by 2. $O_3$, satisfies one query at Cell $(4,4)$, and hence the aggregates of Row~4 and Column~4 are incremented. The overall cost of the evaluator gets the value of 4. Maintaining these aggregates requires $O(1)$ processing time per data object. Tornado maintains similar row, column and overall aggregates for the number of queries within grid cells.

To maximize the cost reduction resulting from splitting a partition, say $X$, into $X_1$ and $X_2$, Tornado tries to minimize the value of $|cost(X_1) - cost(X_2)|$ by trying all possible vertical and horizontal splits. If Equation~\ref{eq:partitioncost} is applied directly, it requires $O(m \times n)$ to find the best split. 
Instead, Tornado uses the row and column aggregates to find the best split in $O(m+n)$. In particular, Tornado scans the column aggregates and keeps a sum of the scanned aggregates, say $S_a$. Initially, $S_a = 0$, and keeps accumulating values from the column aggregates as long as $S_a$ is less than half the overall cost of the evaluator, say $(O_{half})$. If $S_a$ is equal to $(O_{half})$, no more aggregates are scanned. If $S_a$ is greater than $(O_{half})$, then the split position is marked, and the same process is repeated, but with the row aggregates. The split position that minimizes the value of $|cost(X_1) - cost(X_2)|$ is chosen.

For example, in Figure~\ref{fig:finegridcost}(b), the best vertical split is between Columns~3 and~4, with a difference of~3 in cost. However, the best horizontal split is between Rows~2 and~3, with a difference of~0 in cost. Hence, the horizontal split is chosen. 

For the shift operation, we need to distinguish between a corner shift and a horizontal/vertical shift. In the corner shift in Figure~\ref{fig:shiftcases}(c), there are no multiple choices for the shift coordinate in $A$. The corner shift coordinate depends on the position of $B$ relative to $A$. This allows $A$ to identify the cost of the cells involved in any shift operation as well as the cell transfer overhead. Notice that there are at most 8 possible corner shifts for any given evaluator.  However, there is no fixed coordinate for the horizontal/vertical shift in $A$. 
The reason is that 
the optimal
coordinate for 
a
horizontal/vertical shift 
depends on the cost of $B$ 
that 
is unknown to $A$. To address this issue, Tornado delays the choice of the best shift coordinate in $A$ until the routing unit makes a decision to perform a horizontal/vertical shift. 

At the time when the routing unit makes a decision as to whether to re-balance or not, it has accurate statistics for both the split/merge and the corner shift operations. The routing unit does not know the exact cost reduction and cell transfer overhead of horizontal/vertical shift operations. The routing unit estimates that an optimal horizontal/vertical shift from evaluator $A$ to evaluator $B$  results in an optimal division of workload between $A$ and $B$. Thus, the estimated cost reduction is computed as $cost(A)-\frac{cost(A)+cost(B)}{2}$.
Assuming uniform query distribution in $A$, the routing unit estimates the cell transfer overhead to be proportional to the amount of workload transferred, i.e., $\beta \times queryCount(A)\times \frac{cost(A)-\frac{cost(A)+cost(B)}{2}}{cost(A)}$. Then, the routing unit chooses the re-balancing operation if necessary. If the re-balancing operation is a horizontal/vertical shift, 
then
the routing unit informs the evaluators involved in this horizontal/vertical shift operation with the costs necessary to make an optimal shift operation similar to finding the optimal split described previously.

\subsection{Updating the Data Structures}\label{subsec:updates}
\noindent
\textbf{Updates to the A-Grid}
Upon a split/merge or a shift operation, Tornado incrementally updates the A-Grid structure described in Section~\ref{sec:tornadostructure}. In this updates, the boundaries of partition in the partitions map (PM) is updated according to the changes in partitioning. Also, A-Grid cells belonging to new partitions, have their partition ids updated.



\noindent
\textbf{Updates to the Textual Summaries}
Upon a split/merge or a shift operation, some queries are transferred from one evaluator to another. For example, consider a shift operation from Evaluator $X$ to another one say, Evaluator $Z$. The routing units are not aware of which queries are transferred. To ensure correct execution and to avoid missing output results, in the routing units, the entire textual summary of $X$ is copied to the textual summary of  $Z$. This may result in having keywords in the textual summary of $Z$ that do not correspond to any query in $Z$. We discuss how to remove those extra keywords in Section~\ref{subsec:falsepositives}.

\noindent
\textbf{Updates to Statistics within Evaluators}
Upon a split/merge or a shift operation, some grid cells move from one evaluator to another. This affects the row, column, and overall aggregates stored at the evaluators.
For example, consider a shift operation from Evaluator $X$ to Evaluator $Y$.
For each grid cell, say $C_x$, in Evaluator $X$, we subtract the value $cost(C_x)$ from the overall cost of Evaluator $X$, and from the row and column aggregates containing Cell $C$. For each grid cell, say $C_y$, in Evaluator $Y$, we add the value of $cost(C_y)$ to the overall cost of Evaluator $Y$, and to the row and the column aggregates containing Cell $C$.

\subsection{Correctness during Load-balancing}

A rebalancing operation affects both the routing and the evaluation layers. In the routing layer, the partitioning of the evaluators changes according to the rebalancing operation. In the evaluators, grid cells move from one evaluator, say $E_1$, to another evaluator, say $E_2$. It is challenging to guarantee the correctness during the re-balancing process because data objects and queries arrive during re-balancing, and Tornado cannot afford to halt the processing until the entire re-balancing is done.

An important question to address is \textit{Which evaluator should receive the incoming data objects and queries during the transient phase? }  $E_1$, or $E_2$, or both?
Tornado splits the transient phase into two steps. 
In every step, we define a set of rules that guarantee correct processing in that phase.
The steps of the transient phase are: (1)~\textit{Cell transfer phase} during which index cells are moved across evaluators, and (2)~\textit{Routing unit update phase} during which routing units update their partitioning. 
\begin{figure}
 		\centering
                \subfigure[Before re-balancing.]{	\includegraphics[width=1.6in]	{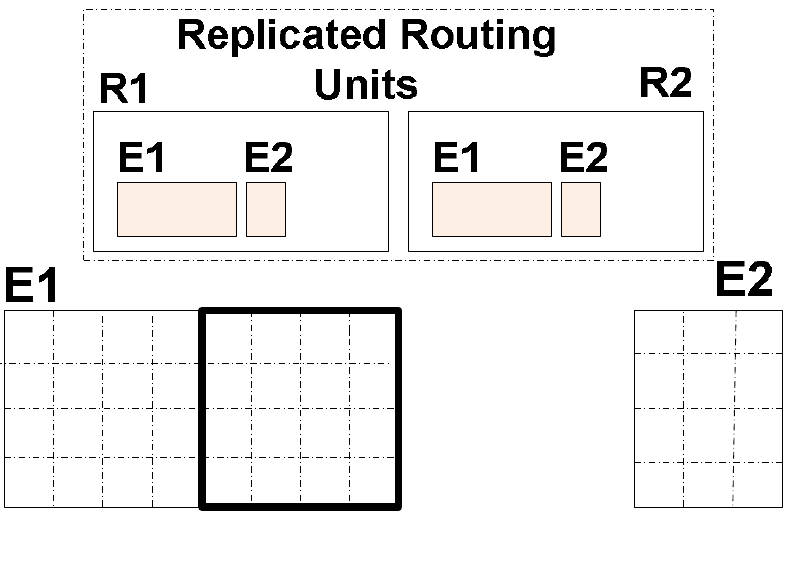}}
        \subfigure[Cell transfer phase.]{	\includegraphics[width=1.6in]	{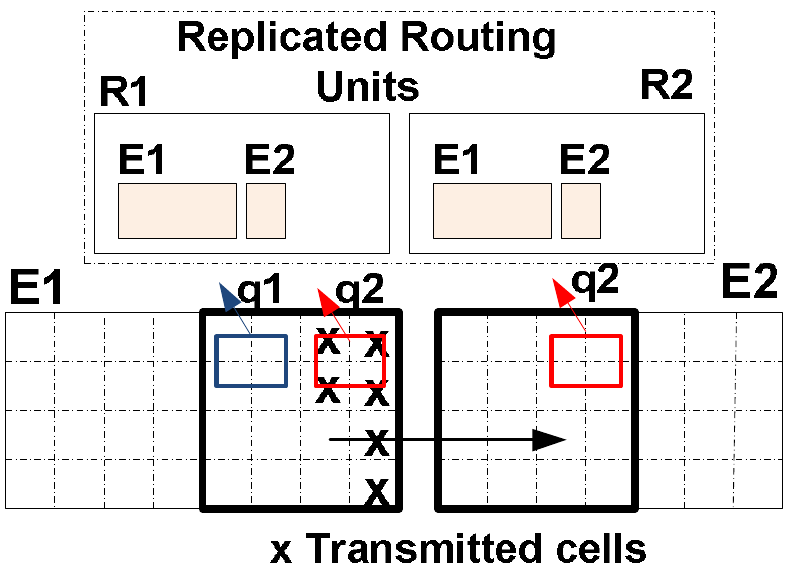}}
        \subfigure[Routing update phase.]{	\includegraphics[width=1.6in]	{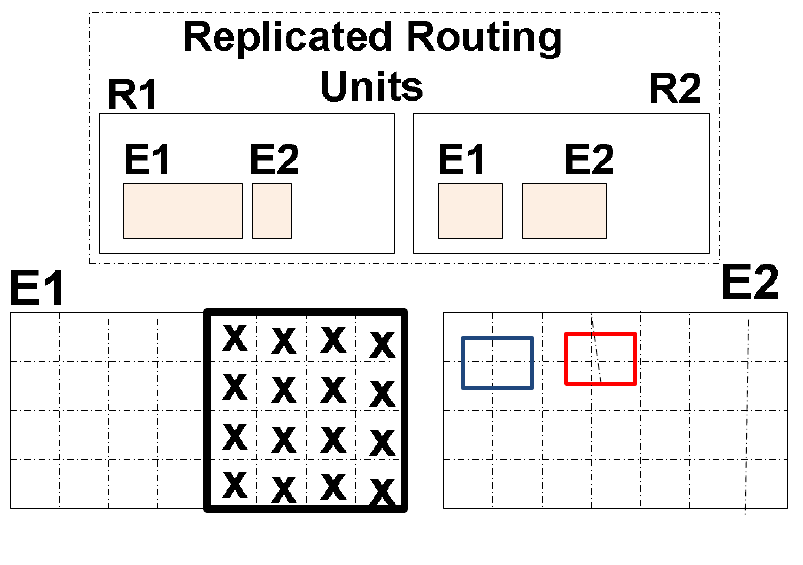}}  
         \subfigure[After re-balancing.]{	\includegraphics[width=1.6in]	{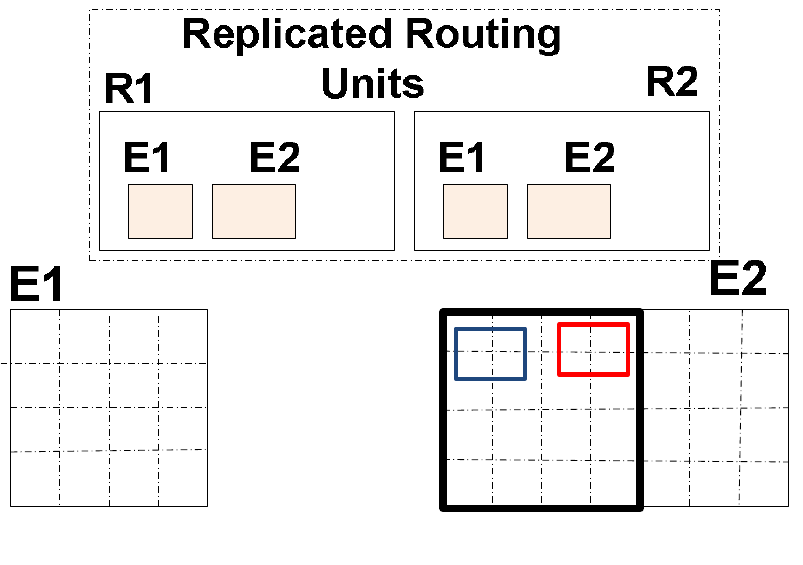}}
	  \caption{Correctness during adaptivity.
     }\label{fig:correctnessphases}
\end{figure}

\noindent
Figure~\ref{fig:correctnessphases} gives an example of a shift rebalancing operation from Evaluator $E_1$ to Evaluator $E_2$. Figure~\ref{fig:correctnessphases}(a) gives the partitioning within routing units and index cells within evaluators before the shift operation. Figure~\ref{fig:correctnessphases}(b) illustrates the \textit{cell transfer phase} from Evaluator $E_1$ to Evaluator $E_2$. During this phase, index cells to be shifted can be in any of two states, namely: \textit{transmitted} and \textit{untransmitted}. A \textit{transmitted} index cell has been moved from Evaluator $E_1$ to Evaluator $E_2$. An \textit{untransmitted} cell is a cell belonging to the partition being shifted and yet to be moved to the destination evaluator. Figure~\ref{fig:correctnessphases}(b) gives the marked transmitted cells.

\noindent
~\textbf{Processing during the cell transfer phase} During the cell transfer phase, all incoming data and queries will be routed to $E_1$ because all routing units use the partitioning before re-balancing. Incoming queries to the area to be shifted are processed according to the following steps:
\begin{enumerate}[noitemsep,nolistsep,leftmargin=*]
\item All incoming queries are processed and indexed in $E_1$
\item If a query arrives to a transmitted cell, forward the query to $E_2$
\end{enumerate}
Incoming data objects are processed in Evaluator $E_1$.

For example, in Figure~\ref{fig:correctnessphases}(b), Evaluator $E_1$ receives Query $q_2$ and stores $q_2$ in the transmitted cells. Then, $E_1$ forwards  $q_2$ to $E_2$. However, query $q_1$ is not forwarded as it arrives to an untrasmitted cell. Notice that, in Figure~\ref{fig:correctnessphases}(b), all routing units have the old partitioning of evaluators.
Queries that overlap untransmitted cells in  $E_1$ will be indexed only in $E_1$, as these cells will shortly be transmitted to $E_2$. 
This guarantees that $E_2$ will eventually receive all queries that arrive during the cell transfer phase. Also, since all data objects that arrive to the shifted cells are evaluated only in $E_1$, then there are no duplicate results.

\noindent
~\textbf{Processing during the routing  update phase}, Due to network delays, it is not possible that all routing units update their partitioning instantaneously. This means that even after the \textit{cell transfer phase}, some routing units may send data and queries to $E_1$ while others send data and queries to $E_2$. 

To address this issue, we adopt the following approach during the routing update phase: \textit{any data object or query that is routed to a \textbf{shifted area} in $E_1$ is neither processed nor indexed in $E_1$ and is instantaneously forwarded to $E_2$}. In Evaluator $E_1$, shifted cells are marked as transmitted.

Figure~\ref{fig:correctnessphases}(c) gives an example to the processing during the routing update phase. Notice that, in Figure~\ref{fig:correctnessphases}(c), the Routing unit $R_1$ has the old partitioning and the Routing unit $R_2$ has the new partitioning. If the Routing unit $R_1$ sends data objects or queries to the shifted cells in $E_1$, then $E_1$ forwards these data objects and queries to $E_2$. This guarantees that there will be no duplicates or missing results as all output during the routing update phase comes from Evaluator $E_2$. Figure~\ref{fig:correctnessphases}(d) gives the routing units and evaluators after the re-balancing shift operation.


\subsection{Lazy Cleaning}\label{subsec:falsepositives}
Queries get dropped and evaluators change boundaries during re-balancing operations. The textual summary at evaluator units needs to be updated to reflect the changes in the keywords of queries within evaluators. Having an outdated textual summary will result in many false positives, and hence affecting the overall system performance. Instead of eagerly updating the textual summary whenever a query is removed, we use a lazy textual summary update approach. In this approach, evaluators periodically send textual summaries to routing units. Evaluators calculate their textual summary in a lazy manner. A background garbage cleaning process visits all fine grid cells, builds the textual summary as cells get visited. It also removes all expired queries with the cell being visited. When a complete cleaning cycle has visited all cells, the textual summary is sent to the evaluators. This approach reduces the overall overhead for textual summary update overhead. 

\section{Analysis}\label{sec:analysis}

In this section, we formally analyze how to set the granularity of the A-Grid. When setting the granularity of the grid, we need to consider both the query registration overhead and the data processing overhead within evaluators. Let $\lambda _d,\lambda _q$ be the arrivals rates of data and queries, respectively. Assume that the average number of queries registered in the system is fixed.
That is on average, the rate of query arrival is equal to the rate of query expiration. 
For example, assume that, on average, at any point in time, there are $k$ continuous queries registered in the system.
To simplify the analysis, assume that we have square queries. Let $r_q$ be the average query side-length. Let $r_c$ be the grid cell side length.  Assume further that we have a unit side-length for the entire space. The total number of evaluators $\rho _e$ can be calculated as follows:
\begin{equation}
{ \rho _e= \lambda _d \times F( \delta )+ \lambda _q \times \gamma}
\end{equation}
where  $\delta $ is the average number of queries per grid cell,  $\gamma$ is the average number of cells a single query overlaps, and $F$ is a function defining the average number of queries relevant to a data object within a grid cell. $F( \delta )$ represents the average processing time of a data object within an evaluator. We aim to minimize the total number of evaluators needed, i.e., $\rho _e$. 

When the average query side-length is less than the grid cell side-length, i.e.,  $r_q < r_c$, a single query can overlap at most four grid cells. That is $\gamma$=$c$, where $c$ is a constant less than $4$. Assuming uniform data and query distribution across the space, the number of queries per cell $\delta$ can be calculated as follows:
\begin{equation}
 \begin{split}
\delta &= c\times \dfrac{k}{total\ number\ of\ grid\ cells } \\
&= c\times \dfrac{k}{\frac{entire\ space}{space\ of\ a\ single\ cell} } \\
Assuming&\ unit\ space\\
       &= c\times   \dfrac{k} {\frac{1}{(r_c)^2}}= c \times  k \times (r_c)^2
\end{split}
\end{equation}

Hence, the total computational overhead when $r_q < r_c$ is:
 \begin{equation}\label{eq:grideq1}
{ \rho _e= \lambda _d \times F(c\times k \times (r_c)^2)+ \lambda _q *c}
\end{equation}
The smaller the grid cell side-length, i.e., $r_c$, the smaller the number of evaluators needed. This can be tempting to use a very small grid cell side-length. However, the smaller the grid cell side-length, the higher the number of cells overlapping
a query. 

In other words,
when the average query side-length is longer than the grid cell side-length, i.e., when $r_q \geq r_c$, the average number of cells per query  $\gamma$ can be calculated as follows:
 \begin{equation}
 \gamma=\frac{ave\ space\ of\ a\ query}{space\ of\ a\ cell}=\frac{(r_q)^2}{(r_c)^2}
 \end{equation}
and the average number of queries per cell is 
 \begin{equation}
 \begin{split}
\delta &= \gamma\times \dfrac{k}{total\ number\ of\ grid\ cells }=\gamma\times k \times   (r_c)^2 \\
       &=(\frac{r_q}{r_c})^2\times k\times (r_c)^2=k\times (r_q)^2
\end{split}
\end{equation}
Hence, the total computational overhead when $r_q \geq r_c$ is:
 \begin{equation}
{ \rho _e= \lambda _d \times F( k \times (r_q)^2)+ \lambda _q \times(\frac{r_q}{r_c})^2}
\label{eq:grideq2}
\end{equation}
In Equation~\ref{eq:grideq1}, when $r_q < r_c$,  $\rho _e$ decreases as $r_c$ decreases. 
In Equation~\ref{eq:grideq2}, when $r_q \geq r_c$, $\rho _e$ decreases as $r_c$ increases. 
In conclusion, 
in order to minimize the number of evaluators needed, we set $r_c=r_q$, i.e.,  set the side-length of the grid cell to be equal to the average query side-length.

\begin{figure*}[t!]
\centering
\begin{minipage}{.4\textwidth}
\raggedleft
        \subfigure[Throughput.]{	\includegraphics[width=1.3in]{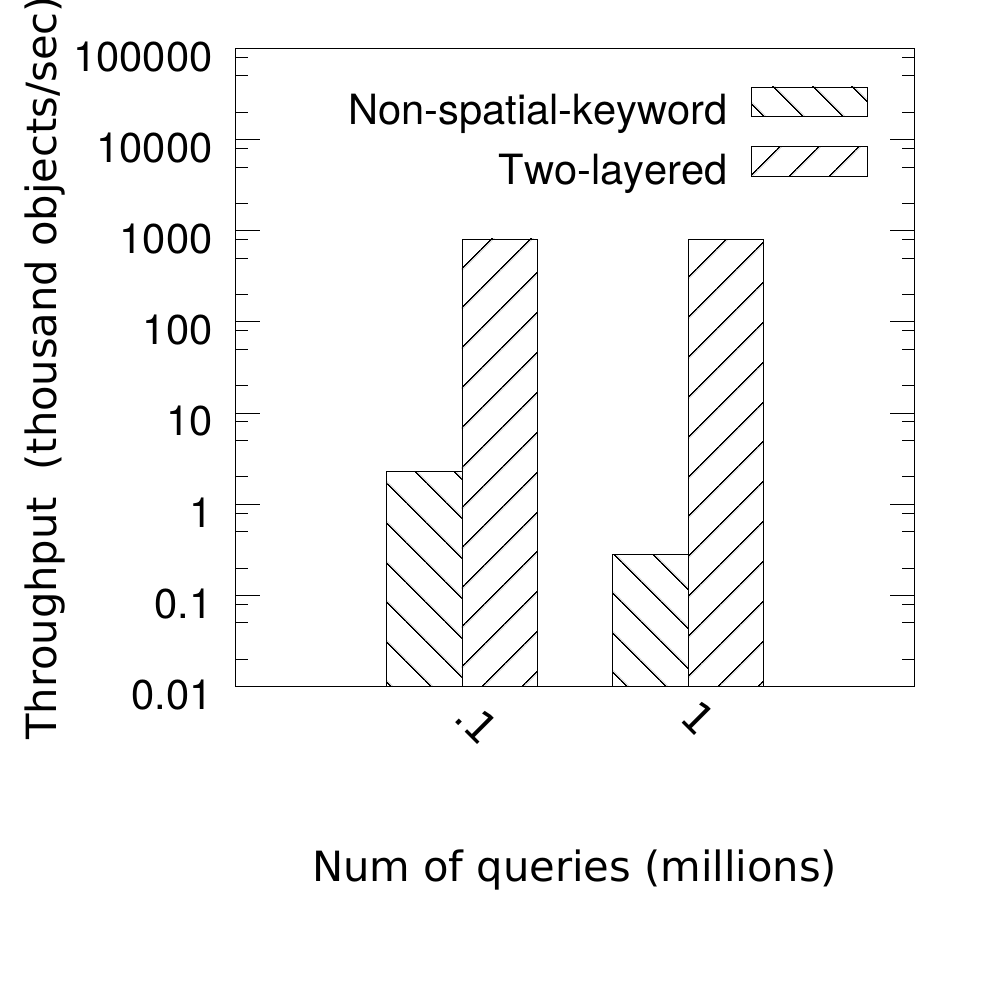}}
        \subfigure[Evaluation latency.]{	\includegraphics[width=1.3in]	{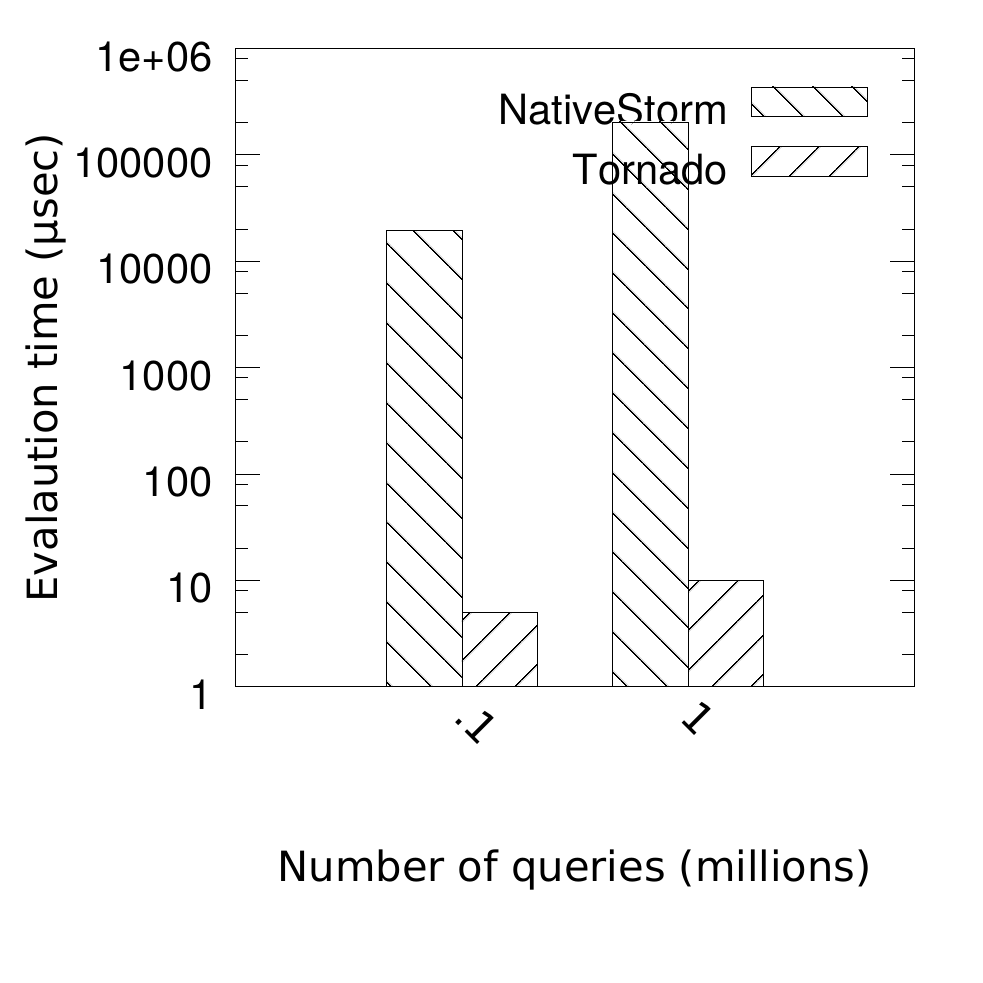}} 
        \caption{Storm native Vs. Tornado.}\label{fig:stormvstornadoscale}
 \end{minipage}%
 \begin{minipage}{.6\textwidth}
 \raggedright
         \subfigure[Throughput.]{	\includegraphics[width=1.3in]	{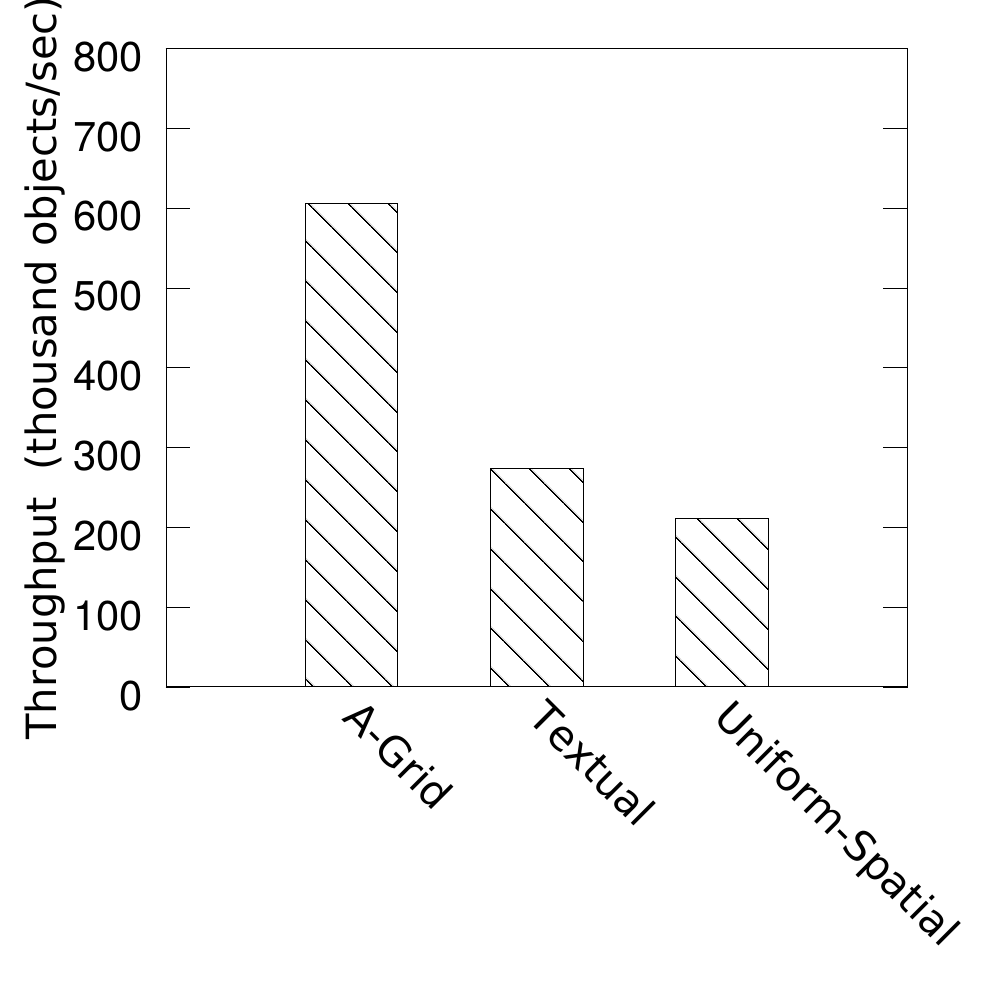}}
        \subfigure[Evaluation latency.]{	\includegraphics[width=1.3in]	{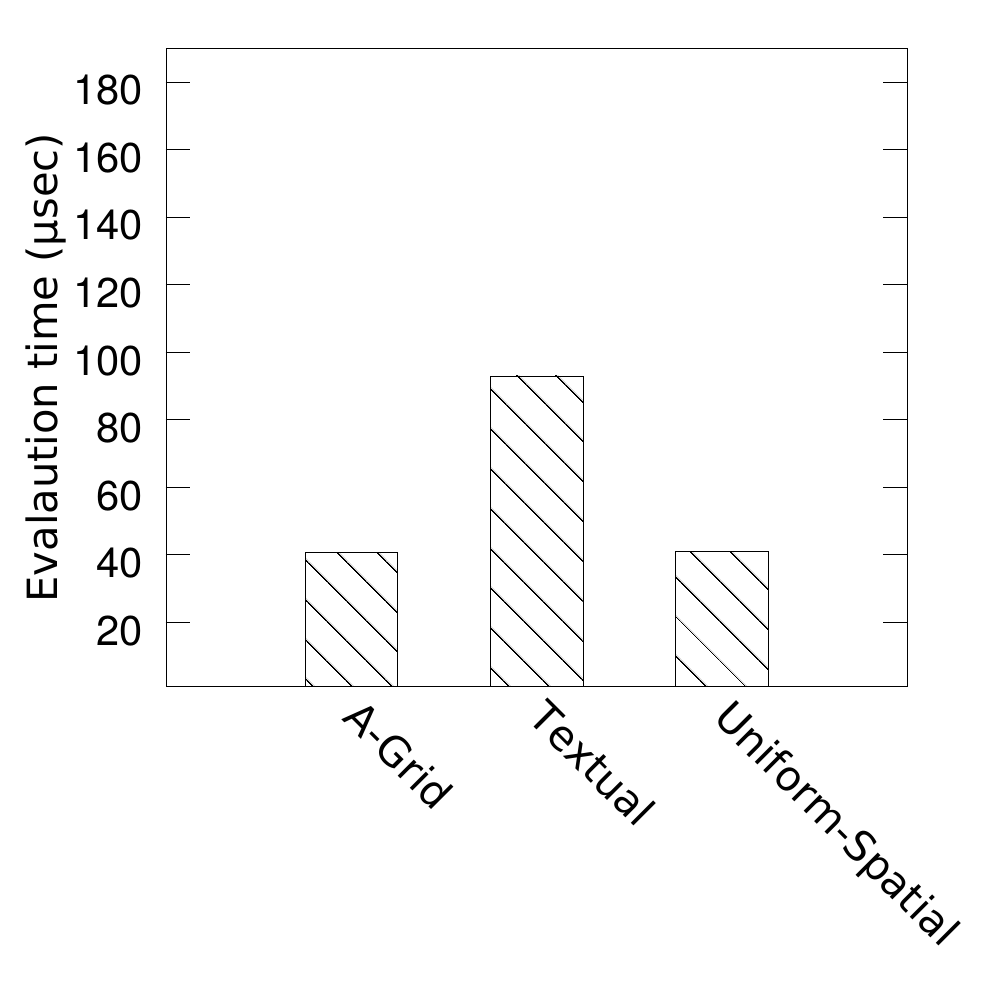}}
        \subfigure[Effect of keyword frequency.]{	\includegraphics[width=1.3in]	{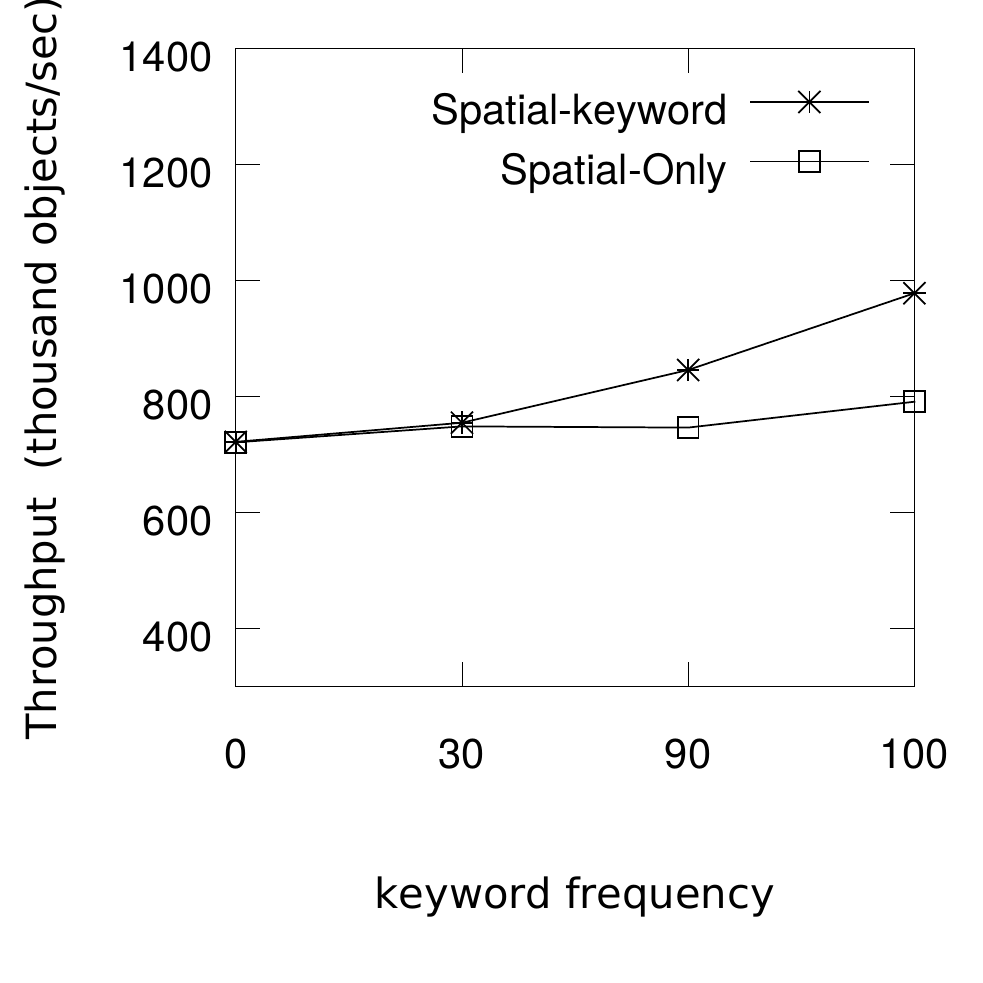}}
        \caption{The performance of routing alternatives.}\label{fig:globallocal}
  \end{minipage}%
\end{figure*}
\begin{table}[!t]
\centering
    \caption{The values of the parameters used in the experimental evaluation.}     
    \label{tab:experiments}

    \begin{small}
    \begin{tabular}{|l|l|l|l|}
    \hline
    {\bfseries Parameter} & {\bfseries Value} \\
     \hline
    Number of evaluators       & 4, 9, 16,~\textbf{25}, 36, 64, 100	\\
    \hline
    Number of routing units       &1, 3, 5,~\textbf{7}, 10, 12 	\\
    \hline
    Number of queries (million)       &1, 2.5, 4,~\textbf{5} 	\\
    \hline
    Number of query keywords       &1, 2,~\textbf{3}, 5, 7	 \\
    \hline
    Spatial side length of a query &.01\%,.05\%,\textbf{.1\%},.5\%,1\%,1.5\%		\\
        \hline
    \end{tabular}
    \end{small} 
\end{table}
\section{Experimental Evaluation}~\label{sec:experimentalevaluation}
In this section, we evaluate the performance of Tornado.
Our experiments are conducted on a 6-node cluster, where each node is a Dell r720xd server that has 16~Intel~E5-2650v2 cores, 64~GB of memory, 48~TB of local storage, and a 40~Gigabit Ethernet interconnect. The cluster runs 20 virtual machines where each virtual machine has 4 cores and 10~GB of memory. Each virtual machine runs Storm 1.0.0 over Centos Linux 6.5. 
We evaluate the performance of Tornado using real datasets and a synthetic query workload. We use a real dataset from Twitter that is composed of 1 billion tweets  with geo-locations inside the US and of size 140 GB. These tweets are collected from January 2014 to March 2015. The format of the tweet, is "id, geo-location, text". We use these tweets to simulate a \textit{continuous} and \textit{infinite} stream of spatio-textual objects such that when all the tweets are streamed, we restart streaming the tweets from the beginning. We use three query datasets each of 5 million tweets, namely; (1)~\textit{normal tweets}, (2)~\textit{spatially skewed}, and (3)~\textit{textually selective}. The \textit{normal tweets} dataset uses the locations and keywords of the tweets as the locations and the keywords of the query. The \textit{spatially-skewed} dataset uses a skewed spatial distribution of tweets. We use the \textit{spatially-skewed} dataset to study the effectiveness of load-balancing techniques. The \textit{textually-selective} dataset sorts keywords of tweets based on their frequencies. Using the frequencies of keywords, we set the textual selectivity of queries. 

Table~\ref{tab:experiments} summarizes the values of the parameters we use. 
We set the default number of query keywords to 3, which resembles the average number of keywords in web searches~\cite{keywordsquery}. The default spatial range of queries is .1\% of the entire spatial range. Following the analysis in Section~\ref{sec:analysis}, we use a 1000$\times$1000 grid. Each experiment runs for 20 minutes to accurately measure the throughput of the system and to avoid any transient behavior.


\begin{figure}[t!]
\centering
    \subfigure[Point routing time]{	\includegraphics[width=1.35in]	{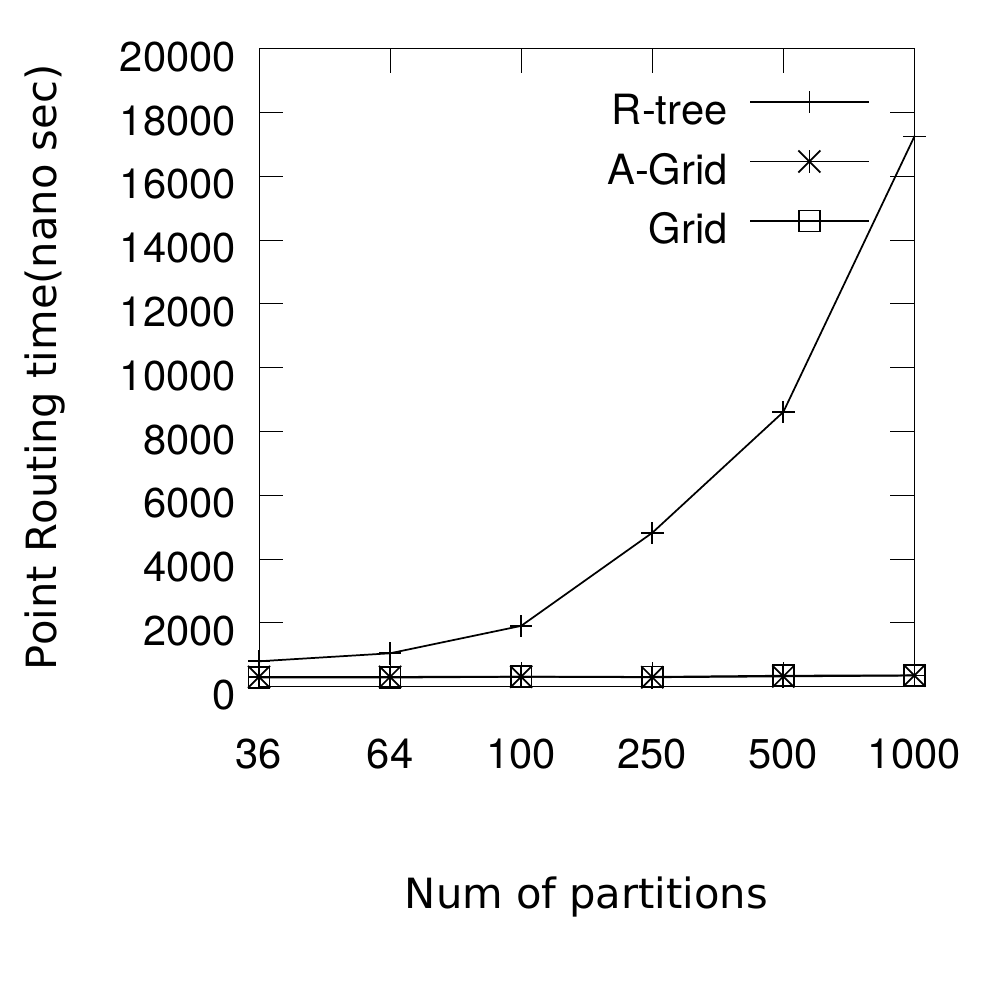}}
        \subfigure[Range routing time]{	\includegraphics[width=1.35in]	{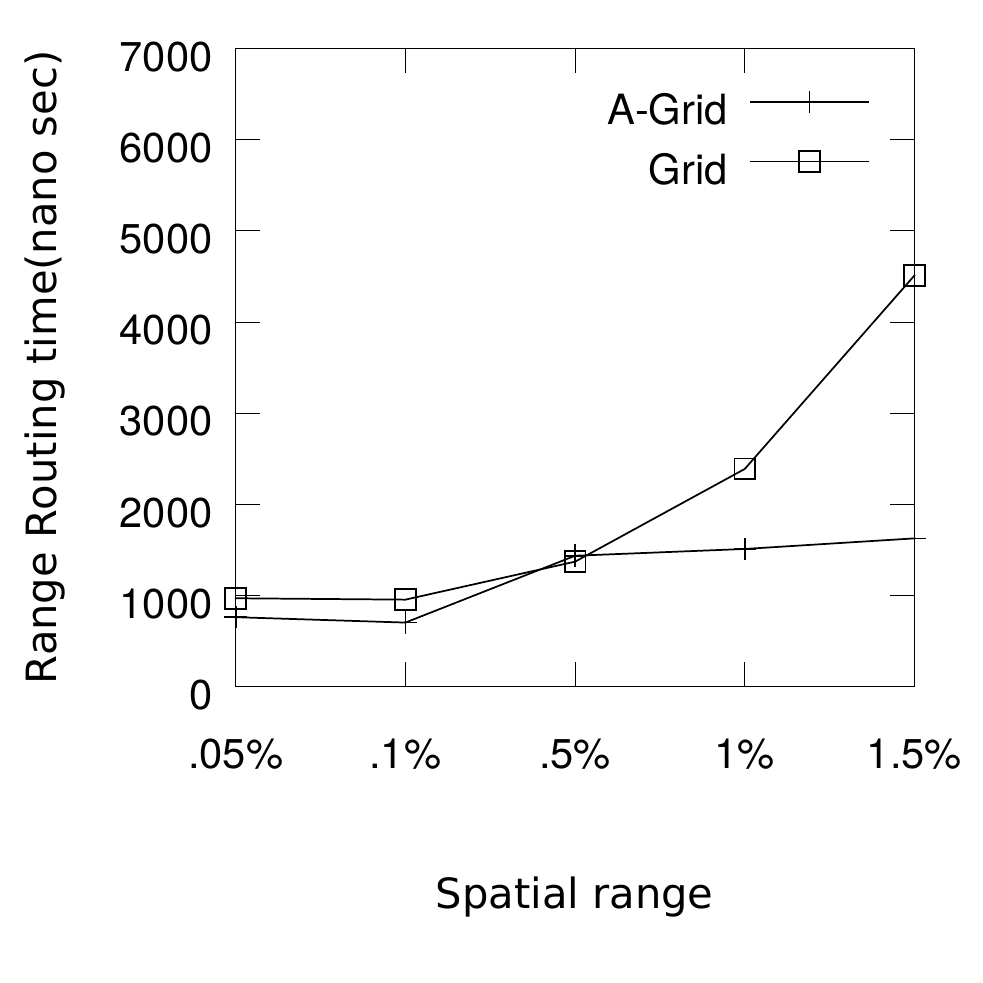}}
        \caption{Spatial routing time for points and ranges.}\label{fig:augmentedgridrouting}

\end{figure}

\subsection{Tornado vs. Native Storm}
In Tornado, the execution is divided between the routing units and the evaluators. However, in a native storm implementation, there is no such distinction. For fairness of evaluation, we set the number of evaluators in the native implementation to be equal to the number of evaluators and routing units in Tornado. In the native storm implementation, all queries are sent to all evaluators, and data objects are randomly distributed across evaluators. In the native approach, a data object needs to be checked against all queries with in an evaluator. This creates a substantial overhead. Figure~\ref{fig:stormvstornadoscale} demonstrates that Tornado achieves more than two orders of magnitude improvement in the overall system throughput and query latency. 

\subsection{Performance of the Routing Layer}
In this experiment, we measure the performance of the following routing alternatives: (1)~the spatial-keyword \textbf{A-Grid}, (2)~\textbf{Textual routing} where the keywords of data objects and queries are used to hash and route data objects and queries to evaluators, and (3)~\textbf{Uniform} spatial partitioning is used in the routing units, where evaluators span equal and non-overlapping spatial ranges regardless of the distribution of workload.

Figures~\ref{fig:globallocal} (a) and (b) show that using the A-Grid in the routing layer achieves the highest throughput and the least processing latency. Using uniform spatial partitioning in the routing layer results in a throughput that is 2 times lower than that of A-Grid. The reason is that using uniform spatial partitioning does not account for the skewed nature of data objects and queries and results in unfair workload distribution across evaluators. Figures~\ref{fig:globallocal} (a) and (b) also illustrate that using textual partitioning of data and queries results in a throughput that is 2 times lower than that of the A-Grid. The reason is that data objects typically have multiple keywords. Textual partitioning replicates data objects to multiple evaluators. This creates a bottleneck in the network bandwidth and reduces the overall throughput and results in having an evaluator processing more data objects.

Figure~\ref{fig:globallocal}(c) demonstrates the effectiveness of spatial-keyword routing against spatial-only routing. In this experiment, we vary the frequency of query keywords from 0\%, i.e., least frequent keywords that do not match the keywords of data objects, to 100\%, i.e., most frequent keywords.
Figure~\ref{fig:globallocal}(c) illustrates that, as the frequency of query keywords decreases, the overall system throughput increases. The reason is that, as the frequency of query keyword decreases, the number of data objects with keywords overlapping with the textual summaries in the A-Grid decreases. This results in having fewer data objects being forwarded to evaluators and hence a reduction of both the computational overhead in the evaluators and the communication overhead between the routing units and the evaluators.

In Figure~\ref{fig:augmentedgridrouting}, we contrast the performance of the A-Grid against the performance of traditional spatial indexes in the routing layer. We study the following alternatives:1)~Standard \textbf{R-tree}, 2)~\textbf{A-Grid}, and 3)~Uniform \textbf{Grid}. Figure~\ref{fig:augmentedgridrouting}(a)  gives the routing times for data points while increasing the number of partitions. As the number of partitions increases, the routing time of the data points increases for the R-tree, and remains constant for both the Grid and the A-Grid. Hence using the R-tree as the routing index is inefficient especially with a high number of evaluators. Although the Grid and the A-Grid have similar performance for point routing, Figure~\ref{fig:augmentedgridrouting}(b) shows that the A-Grid outperforms the Grid for range routing.  
We increase the spatial range of queries from .05\% to 1.5\% of the entire spatial range. This is due to the effectiveness of the neighbor-based routing algorithm used in the A-Grid. We conclude that the A-Grid achieves the least routing time for both points and ranges.

\begin{figure}[t!]
      \centering
      \subfigure[Data objects throughput]{
      \includegraphics[width=1.35in]{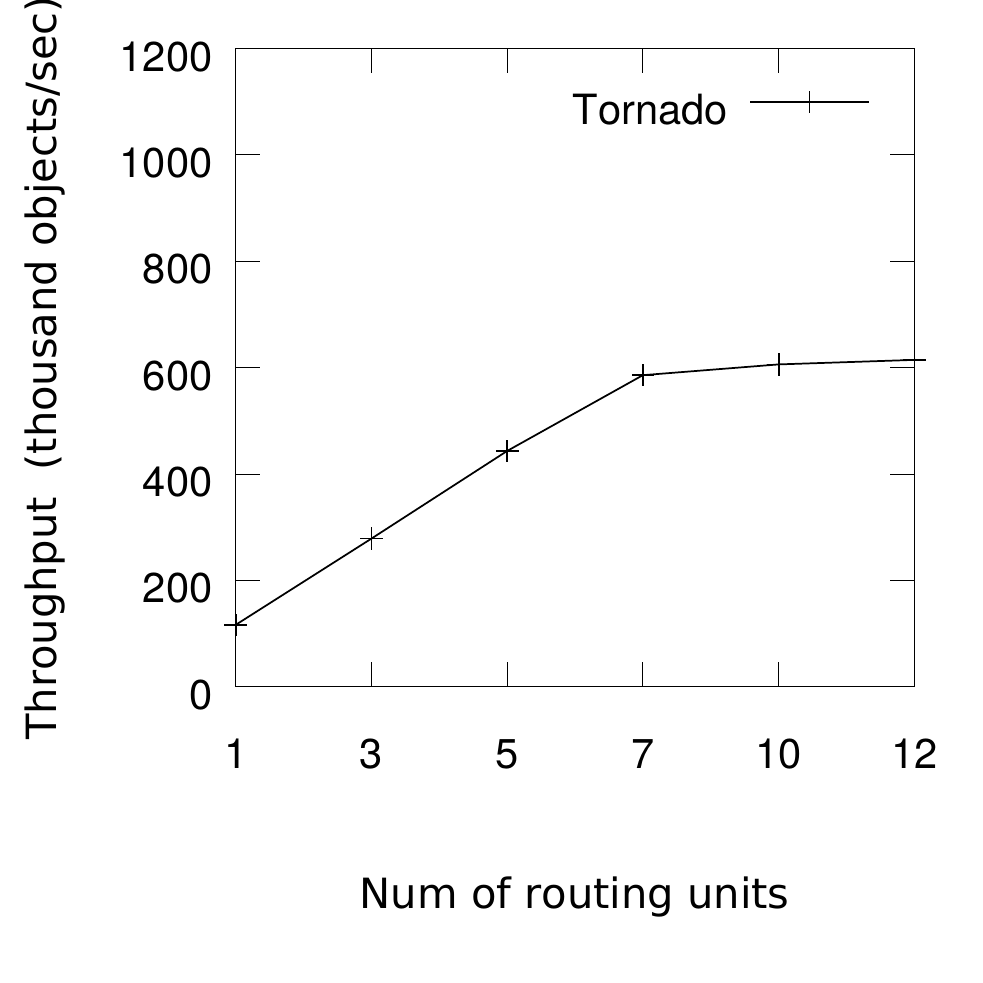}}
         \subfigure[Query throughput.]{
      \includegraphics[width=1.35in]{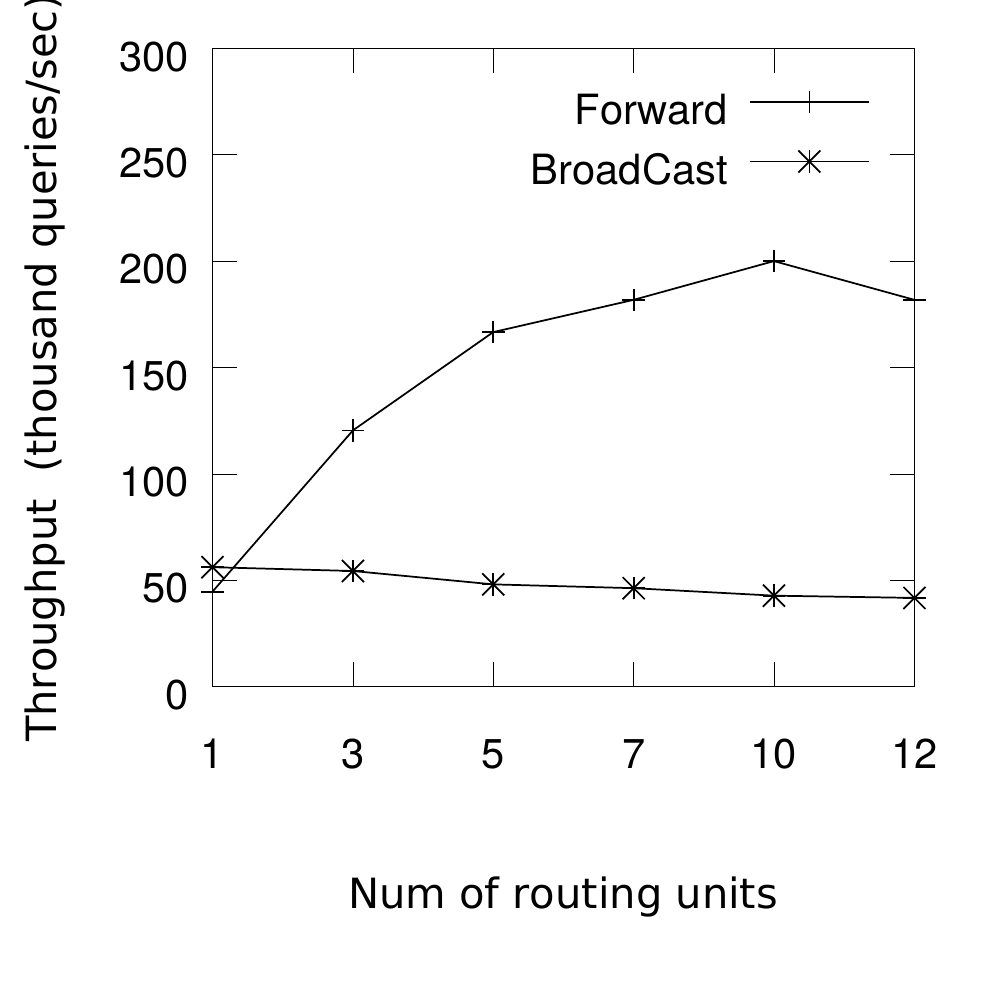}}
       \caption{Number of routing units effect.}\label{fig:numberOfrouting}
\end{figure}

In Figure~\ref{fig:numberOfrouting}, we study the effect of the number of routing units on the overall system throughput. Figure~\ref{fig:numberOfrouting}(a) gives the throughput when increasing the number of routing units.  If there is only one routing instance, then the routing layer becomes a bottleneck. As we increase the number of routing instances, the system throughput increases. The increase in throughput saturates after 7 routing instances. After that, the bottleneck moves from the routing layer to the evaluation layer. This demonstrates that the routing layer is light-weight, and that we do not need many routing units to forward data objects to evaluators.

In Figure~\ref{fig:numberOfrouting}(b), we contrast two approaches for textual summary distribution across routing units, namely \textit{broadcast} that sends an incoming query to all the routing units, and \textit{forward} that broadcasts the query keywords only when they do not exist in the textual summary of an evaluator.
As the figure illustrates, using the \textit{forward} improves the query throughput by up to five times.
Observe that in Figure~\ref{fig:numberOfrouting}, the data object throughput is higher than the query throughput. This is both expected and acceptable. 
Typically, in real-life applications that data object updates arrive at a much higher rate than that of queries. The reason for the lower throughput of queries is 
that, in contrast to data objects, queries can be forwarded to multiple evaluators. Also, queries need to be stored and indexed within evaluators. 
               
\begin{figure}[t!]
        \subfigure[Spatial range effect.]{	\includegraphics[width=1.31in]	{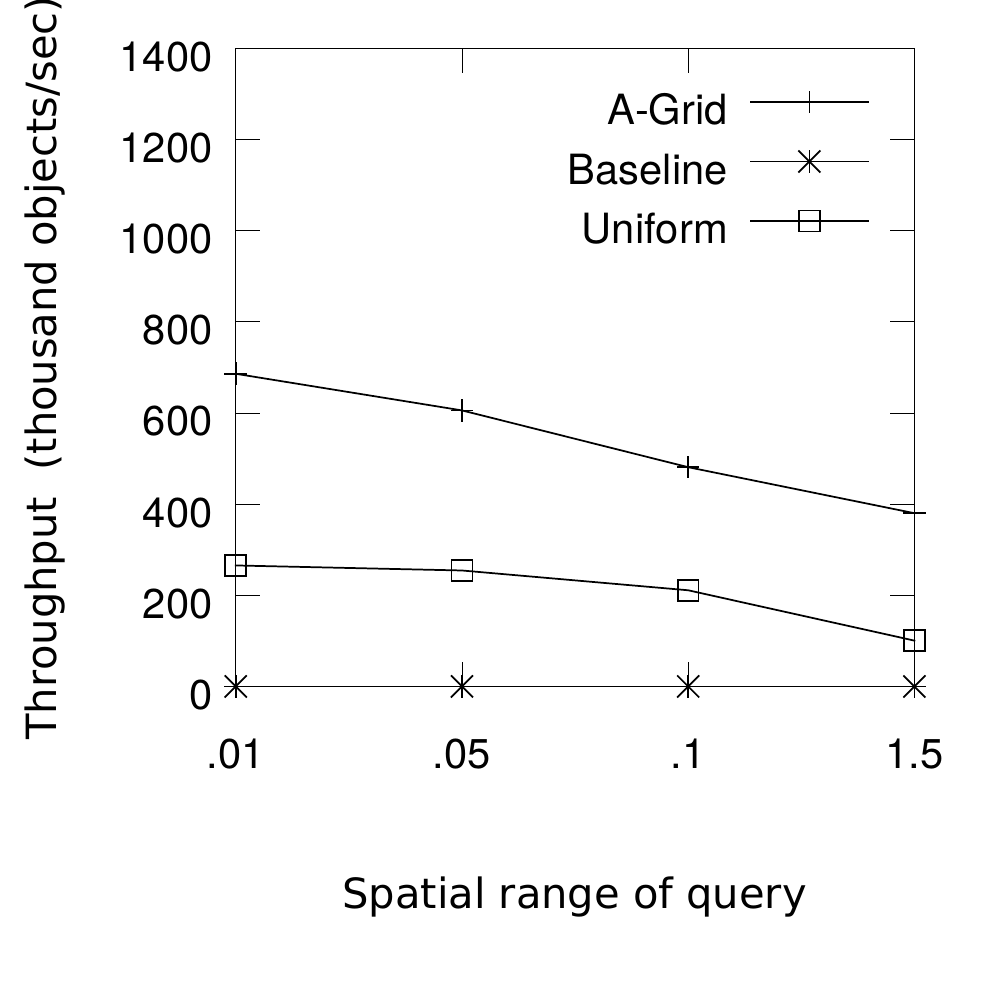}}
        \subfigure[Number of query keywords.]{	\includegraphics[width=1.31in]	{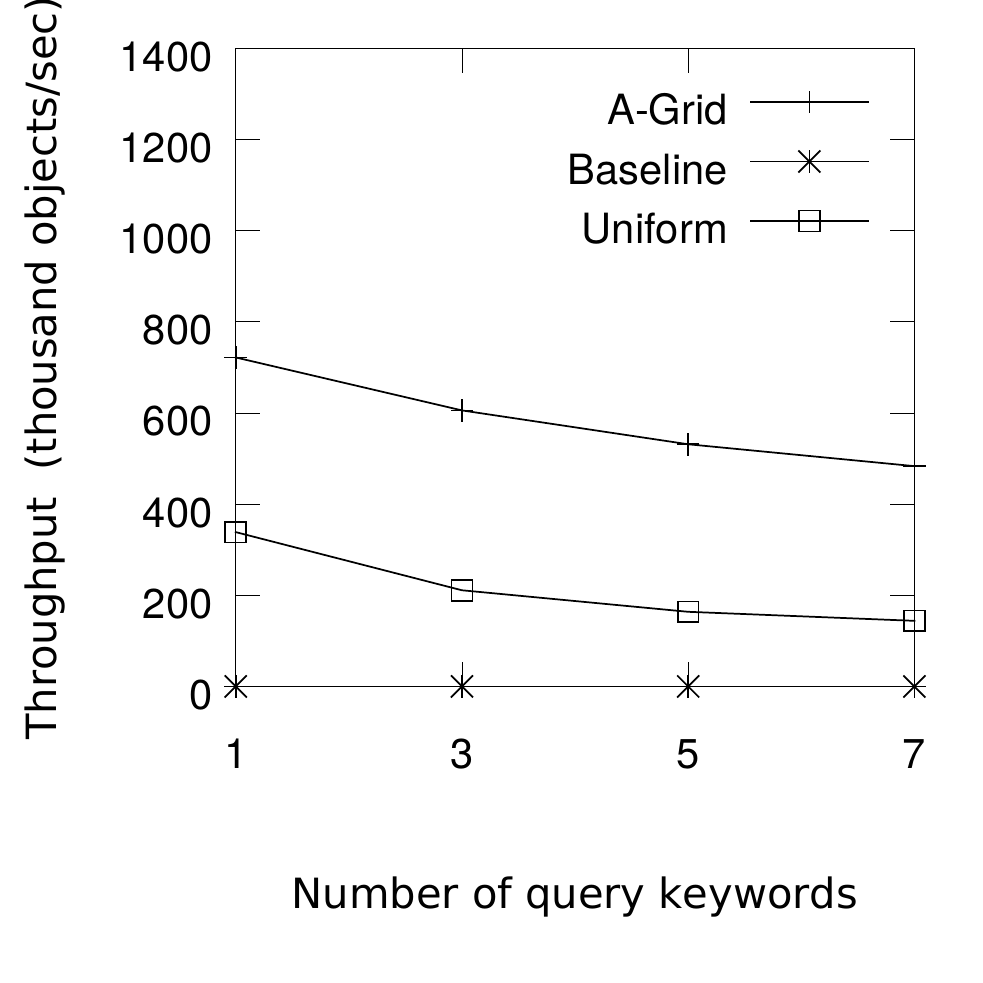}}
         \caption{Performance under various  query workloads.}\label{fig:numberofqueries}
\end{figure}   
\begin{figure}[t!]
        \subfigure[Scalability.]{	\includegraphics[width=1.31in]	{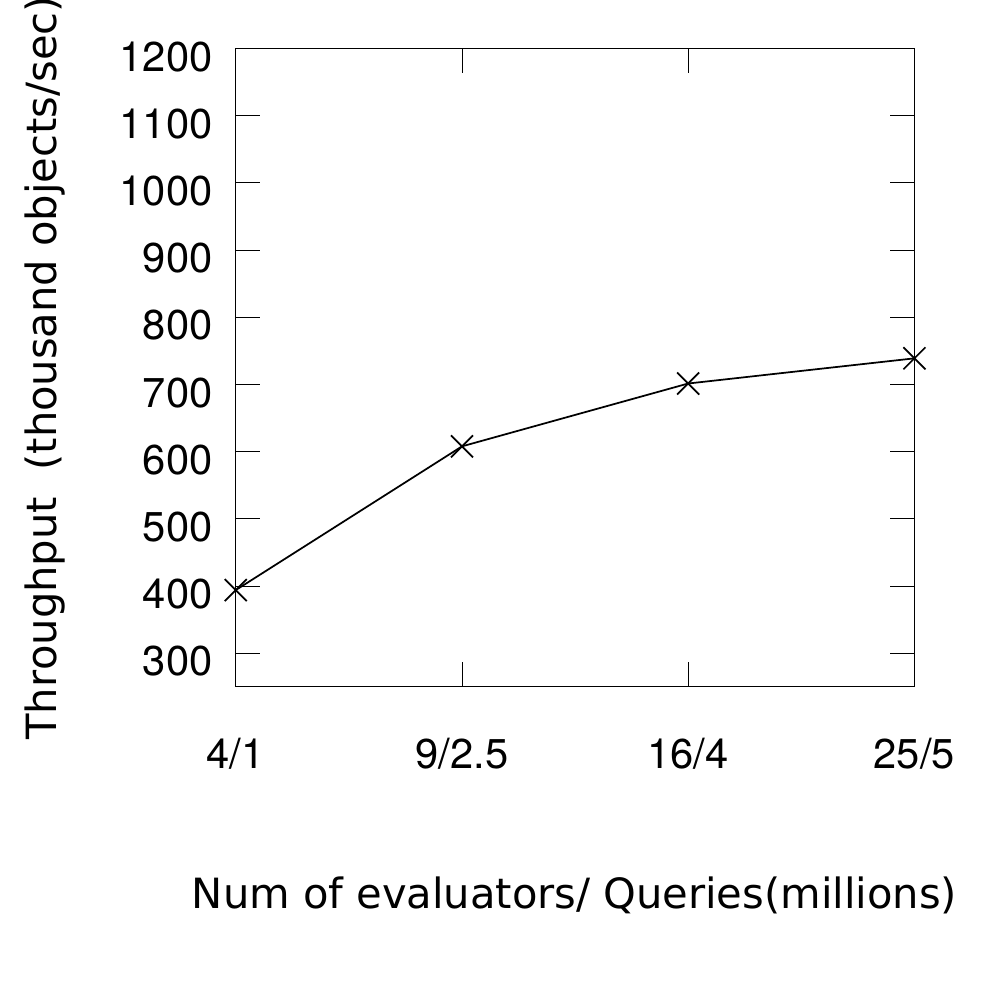}}
        \subfigure[Exceeding cluster resources.]{	\includegraphics[width=1.31in]	{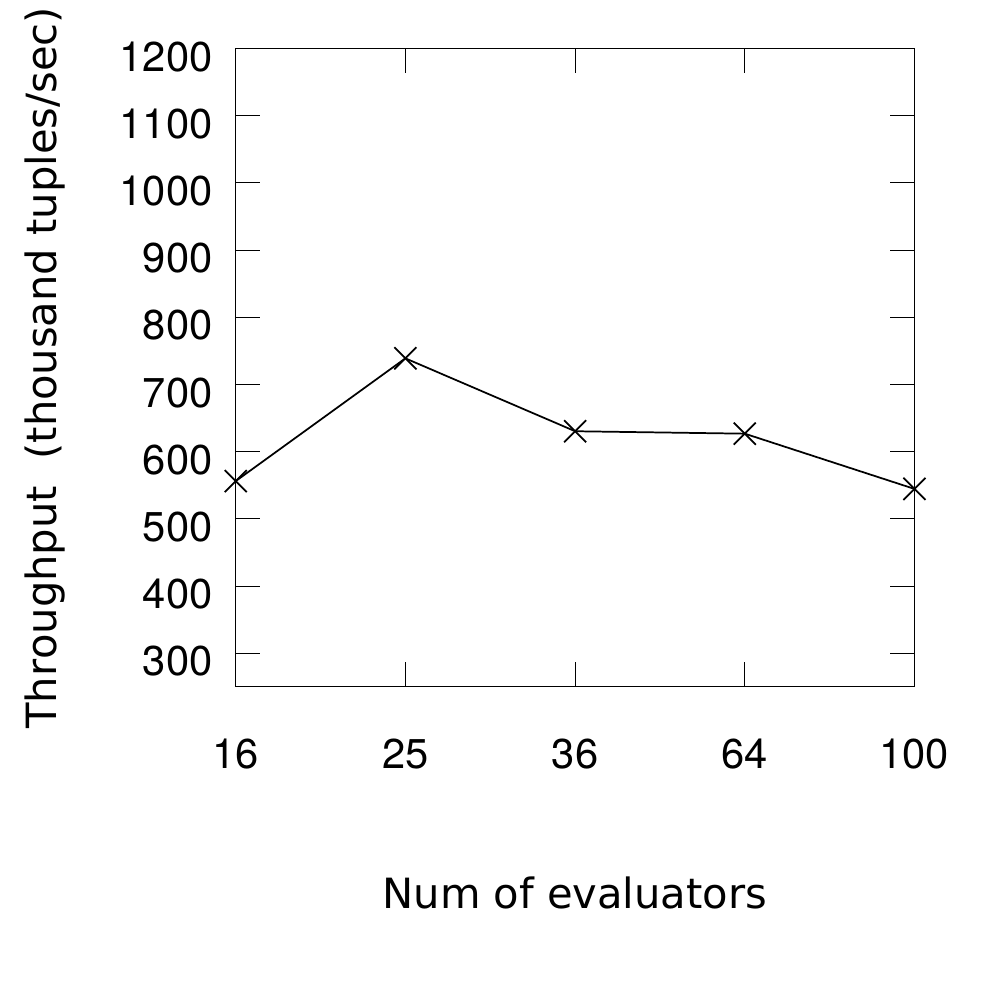}}
         \caption{Scalability against the number of evaluators.}\label{fig:scalability}
  \end{figure}
  \begin{figure}[t!]
        \subfigure[Adaptivity.]{  \includegraphics[width=1.31in]{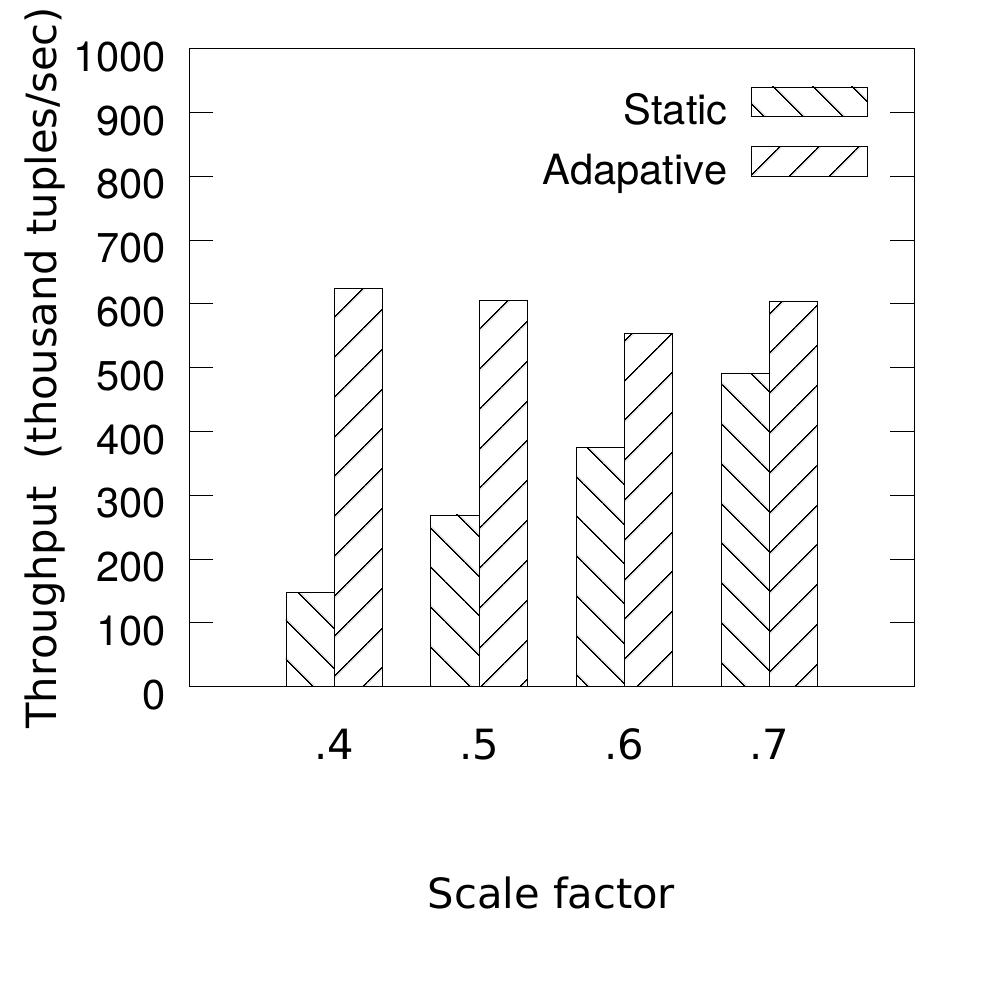}}
         \subfigure[Statistics overhead.]{  \includegraphics[width=1.31in]{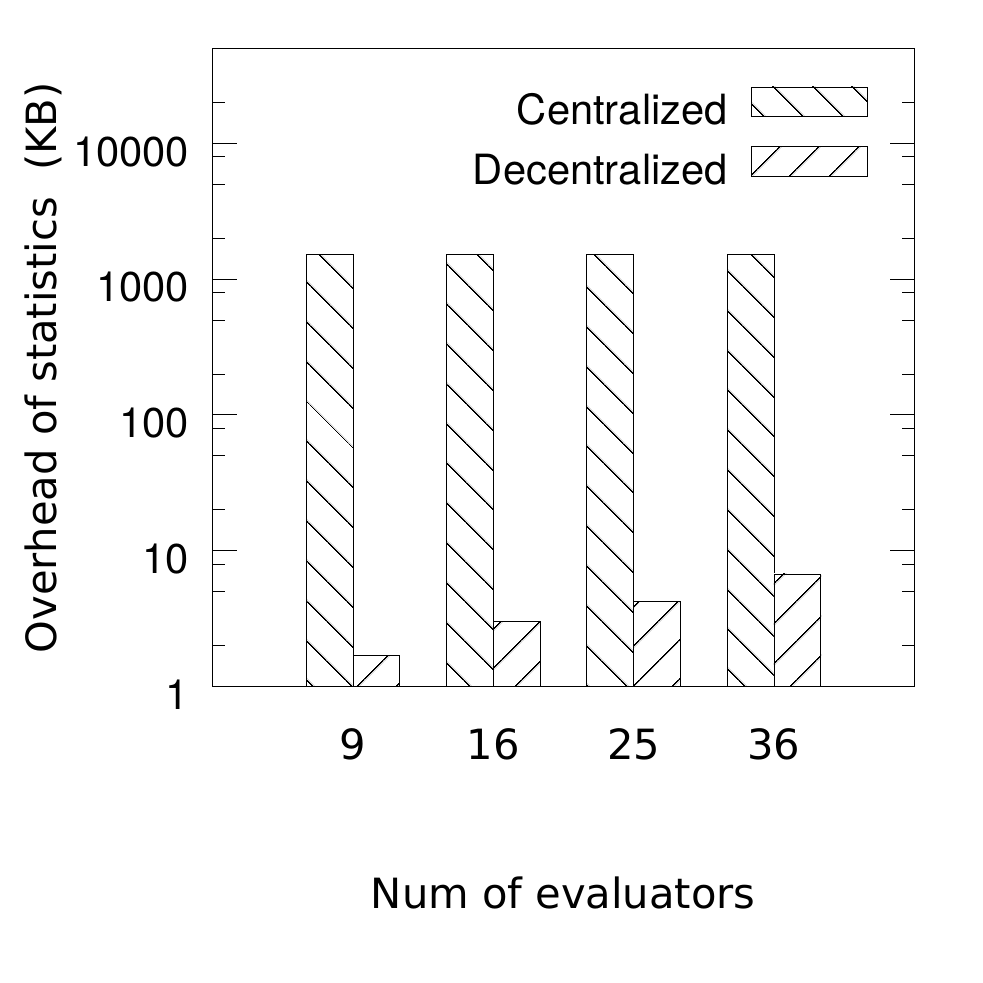}}
        \caption{Adaptivity.}\label{fig:adaptivity}
\end{figure}
\subsection{Scalability}
In this experiment, we study the scalability of Tornado under various query workloads. In Figure~\ref{fig:numberofqueries}(a), we vary the spatial range of the queries from .01\% to 1.5\% of the maximum spatial range. Figure~\ref{fig:numberofqueries}(a) illustrates that Tornado is scalable and that the system throughput is not significantly affected by the increase in the spatial extent of the query.  In Figure~\ref{fig:numberofqueries}(b), we increase the number of query keywords from 1 to 7. Figure~\ref{fig:numberofqueries}(b) illustrates
that Tornado is scalable and that the system throughput is not significantly affect by the increase in the number of query keywords. 
We study the scalability of Tornado while increasing the number of queries and evaluators. In Figure~\ref{fig:numberofqueries}(c), we increase the number of queries and evaluators from 4 evaluators and 1 million queries to 25 evaluators and 5 million queries. Figure~\ref{fig:scalability}(c) illustrates that Tornado scales well when adding more evaluators and queries. However, it is may be tempting to increase the number of evaluators indefinitely. Figure~\ref{fig:scalability}(b) gives the effect of increasing the number of evaluators under a fixed query workload of 5 million queries. Initially, the system throughput increases as we add evaluators. However, given that we only have 20 machines in the cluster, after 25 evaluators, the overall throughput starts to decline due to the contention among the evaluators over the machine resources.

\subsection{Adaptivity}
In this experiment, we compare adaptive partitioning against static partitioning in Tornado. To simulate a change in the workload, we multiply the spatial coordinates, i.e., $x$, and $y$ coordinates, of data objects and queries by a scale factor, say $sf$. This scaling results in directing data objects and queries to only a subset of evaluators. We vary the scale factor $sf$ from .4 to .7. Figure~\ref{fig:adaptivity}(a) illustrates that the adaptive partitioning is able to maintain a stable throughput in contrast to the static partitioning. The smaller the scale factor, the lower the throughput for static partitioning. The reason is that, in the static partitioning, fewer evaluators handle the entire workload. This results in a bottleneck in the evaluation layer. However, in the adaptive partitioning, the routing layer redistributes the workload across evaluators to avoid bottlenecks in the system.

In Figure~\ref{fig:adaptivity}(b), we compare the communication overhead between the decentralized load balancing and a centralized approach. In a centralized load-balancing approach detailed workload statistics need to transmitted to the routing layer. However, in the decentralized load-balancing approach only summaries of statistics are transmitted to the routing layer. Figure~\ref{fig:adaptivity}(b) illustrates that the communication overhead of decentralized load-balancing is much less than the overhead of the centralized load-balancing approach. 

\section{Related Work}\label{sec:relatedwork}
Work related to Tornado can be categorized into four main categories: 1)~distributed query-processing, 2)~spatial and spatio-textual query-processing, 3)~spatio-textual access methods, and 4)~adaptive query-processing.\\
\noindent
\textbf{Distributed Query-Processing:}
Many
systems have been developed to process large-scale datasets. Batch-based systems, e.g., Apache Hadoop~\cite{hadoop}, are designed to process large amounts of data in an offline manner (i.e., on disk). In 
these
systems, a single job can take several minutes or even hours to complete. Apache Spark~\cite{spark} has been introduced to improve the latency of Hadoop. Streaming systems, e.g., Storm\cite{toshniwal2014storm},
process data streams of high arrival rates in real-time. However, none of the aforementioned systems is optimized for processing spatial-keyword queries. 
\\
\noindent
\textbf{Spatial and Spatio-textual Systems:}
Several centralized
systems, e.g.,~\cite{mokbel2008sole} and distributed e.g.,~\cite{aji2013hadoop,eldawy2015spatialhadoop} 
have been proposed to process spatial queries. However, these systems do not support the execution of spatial-keyword queries.
ST-HBase~\cite{ma2013st} is a distributed spatio-textual processing system that is based on HBase. However, ST-HBase is batch-based, i.e., cannot support real-time execution of spatial-keyword queries.
The general range search problem, i.e., the problem of finding the data points and rectangles that overlap a rectangle has been extensively studied in the literature, e.g.,~\cite{agarwal1999geometric,guttman1984r}. Tornado's neighbor-based routing algorithm does not violate the logarithmic lower-bounds described by Agrawal et al.~\cite{agarwal1999geometric} as spatial ranges indexed do not overlap and cover the entire space.\\
\noindent
\textbf{Spatio-Textual Access Methods:}
Many
indexes have been proposed to process spatial-keyword queries e.g.,~\cite{de2008keyword,li2011ir,wang2015ap}. These access methods integrate a spatial index, e.g., the R-tree~\cite{guttman1984r} or the Quad-tree~\cite{samet1990design} with a keyword index, e.g., Inverted lists~\cite{zobel2006inverted}. These access methods are centralized and do not scale across multiple machines.\\
\noindent
\textbf{Adaptive Query-Processing:}
AQWA~\cite{aly2015aqwa} is an adaptive spatial processing system that is based on Hadoop. AQWA executes snapshot queries over static data. AQWA reacts to workload changes by incrementally splitting the data partitions. Unlike Tornado, AQWA requires centralized statistics, and halts the processing of queries until rebalancing is 
completed.
Moreover, AQWA does not consider the textual aspects of the data and the routing cost of the data objects. 
Moreover, 
AQWA uses only split operations to  redistribute the workload.
\section{Conclusions}\label{sec:conculsion}
{\sloppy
In this paper, we introduce Tornado, an adaptive, distributed, and real-time system for the processing spatial-keyword data streams. Tornado uses several optimizations, e.g., spatio-textual global routing, neighbor-based spatial routing, to alleviate performance bottlenecks in the system. Tornado is adaptive to changes in data distribution and query workload and is able to preserve the system throughput under varying workloads. Tornado achieves two orders of magnitude improvements over the performance of the baseline.  }

\bibliographystyle{ACM-Reference-Format}


\begin{thebibliography}{00}


\ifx \showCODEN    \undefined \def \showCODEN     #1{\unskip}     \fi
\ifx \showDOI      \undefined \def \showDOI       #1{{\tt DOI:}\penalty0{#1}\ }
  \fi
\ifx \showISBNx    \undefined \def \showISBNx     #1{\unskip}     \fi
\ifx \showISBNxiii \undefined \def \showISBNxiii  #1{\unskip}     \fi
\ifx \showISSN     \undefined \def \showISSN      #1{\unskip}     \fi
\ifx \showLCCN     \undefined \def \showLCCN      #1{\unskip}     \fi
\ifx \shownote     \undefined \def \shownote      #1{#1}          \fi
\ifx \showarticletitle \undefined \def \showarticletitle #1{#1}   \fi
\ifx \showURL      \undefined \def \showURL       #1{#1}          \fi
\providecommand\bibfield[2]{#2}
\providecommand\bibinfo[2]{#2}
\providecommand\natexlab[1]{#1}
\providecommand\showeprint[2][]{arXiv:#2}

\bibitem[\protect\citeauthoryear{??}{had}{2017}]%
        {hadoop}
 \bibinfo{year}{2017}\natexlab{}.
\newblock \bibinfo{title}{{H}adoop}.
\newblock \bibinfo{howpublished}{\url{http://hadoop.apache.org/}}.
  (\bibinfo{year}{2017}).
\newblock


\bibitem[\protect\citeauthoryear{??}{int}{2017}]%
        {internetstats}
 \bibinfo{year}{2017}\natexlab{}.
\newblock \bibinfo{title}{Internet live stats}.
\newblock \bibinfo{howpublished}{\url{https://internetlivestats.com/}}.
  (\bibinfo{year}{2017}).
\newblock


\bibitem[\protect\citeauthoryear{??}{key}{2017}]%
        {keywordsquery}
 \bibinfo{year}{2017}\natexlab{}.
\newblock \bibinfo{title}{Keyword search statistics}.
\newblock
  \bibinfo{howpublished}{\url{http://www.keyworddiscovery.com/keyword-stats.html}}.
    (\bibinfo{year}{2017}).
\newblock


\bibitem[\protect\citeauthoryear{Agarwal, Erickson, et~al\mbox{.}}{Agarwal
  et~al\mbox{.}}{1999}]%
        {agarwal1999geometric}
\bibfield{author}{\bibinfo{person}{Pankaj~K Agarwal}, \bibinfo{person}{Jeff
  Erickson}, {and} \bibinfo{person}{others}.} \bibinfo{year}{1999}\natexlab{}.
\newblock \showarticletitle{Geometric range searching and its relatives}.
\newblock \bibinfo{journal}{{\it Contemp. Math.}}  \bibinfo{volume}{223}
  (\bibinfo{year}{1999}), \bibinfo{pages}{1--56}.
\newblock


\bibitem[\protect\citeauthoryear{Aji, Wang, Vo, Lee, Liu, Zhang, and Saltz}{Aji
  et~al\mbox{.}}{2013}]%
        {aji2013hadoop}
\bibfield{author}{\bibinfo{person}{Ablimit Aji}, \bibinfo{person}{Fusheng
  Wang}, \bibinfo{person}{Hoang Vo}, \bibinfo{person}{Rubao Lee},
  \bibinfo{person}{Qiaoling Liu}, \bibinfo{person}{Xiaodong Zhang}, {and}
  \bibinfo{person}{Joel Saltz}.} \bibinfo{year}{2013}\natexlab{}.
\newblock \showarticletitle{Hadoop GIS: a high performance spatial data
  warehousing system over mapreduce}.
\newblock \bibinfo{journal}{{\em PVLDB\/}} \bibinfo{volume}{6},
  \bibinfo{number}{11} (\bibinfo{year}{2013}), \bibinfo{pages}{1009--1020}.
\newblock


\bibitem[\protect\citeauthoryear{Aly, Elmeleegy, Qi, and Aref}{Aly
  et~al\mbox{.}}{2016}]%
        {kangroo}
\bibfield{author}{\bibinfo{person}{Ahmed~M Aly}, \bibinfo{person}{Hazem
  Elmeleegy}, \bibinfo{person}{Yan Qi}, {and} \bibinfo{person}{Walid Aref}.}
  \bibinfo{year}{2016}\natexlab{}.
\newblock \showarticletitle{Kangaroo: Workload-Aware Processing of Range Data
  and Range Queries in Hadoop}. In \bibinfo{booktitle}{{\em WSDM}}.
  \bibinfo{pages}{397--406}.
\newblock


\bibitem[\protect\citeauthoryear{Aly, Mahmood, Hassan, Aref, Ouzzani,
  Elmeleegy, and Qadah}{Aly et~al\mbox{.}}{2015}]%
        {aly2015aqwa}
\bibfield{author}{\bibinfo{person}{Ahmed~M Aly}, \bibinfo{person}{Ahmed~R
  Mahmood}, \bibinfo{person}{Mohamed~S Hassan}, \bibinfo{person}{Walid~G Aref},
  \bibinfo{person}{Mourad Ouzzani}, \bibinfo{person}{Hazem Elmeleegy}, {and}
  \bibinfo{person}{Thamir Qadah}.} \bibinfo{year}{2015}\natexlab{}.
\newblock \showarticletitle{AQWA: adaptive query workload aware partitioning of
  big spatial data}.
\newblock \bibinfo{journal}{{\em Proceedings of the VLDB Endowment\/}}
  \bibinfo{volume}{8}, \bibinfo{number}{13} (\bibinfo{year}{2015}),
  \bibinfo{pages}{2062--2073}.
\newblock


\bibitem[\protect\citeauthoryear{Beckmann, Kriegel, Schneider, and
  Seeger}{Beckmann et~al\mbox{.}}{1990}]%
        {beckmann1990r}
\bibfield{author}{\bibinfo{person}{Norbert Beckmann},
  \bibinfo{person}{Hans-Peter Kriegel}, \bibinfo{person}{Ralf Schneider}, {and}
  \bibinfo{person}{Bernhard Seeger}.} \bibinfo{year}{1990}\natexlab{}.
\newblock \bibinfo{booktitle}{{\em The R*-tree: an efficient and robust access
  method for points and rectangles}}. Vol.~\bibinfo{volume}{19}.
\newblock \bibinfo{publisher}{ACM}.
\newblock


\bibitem[\protect\citeauthoryear{De~Berg, Van~Kreveld, Overmars, and
  Schwarzkopf}{De~Berg et~al\mbox{.}}{2000}]%
        {de2000computational}
\bibfield{author}{\bibinfo{person}{Mark De~Berg}, \bibinfo{person}{Marc
  Van~Kreveld}, \bibinfo{person}{Mark Overmars}, {and}
  \bibinfo{person}{Otfried~Cheong Schwarzkopf}.}
  \bibinfo{year}{2000}\natexlab{}.
\newblock \bibinfo{booktitle}{{\em Computational geometry}}.
\newblock \bibinfo{publisher}{Springer}.
\newblock


\bibitem[\protect\citeauthoryear{De~Felipe, Hristidis, and Rishe}{De~Felipe
  et~al\mbox{.}}{2008}]%
        {de2008keyword}
\bibfield{author}{\bibinfo{person}{Ian De~Felipe}, \bibinfo{person}{Vagelis
  Hristidis}, {and} \bibinfo{person}{Naphtali Rishe}.}
  \bibinfo{year}{2008}\natexlab{}.
\newblock \showarticletitle{Keyword search on spatial databases}. In
  \bibinfo{booktitle}{{\em ICDE}}. \bibinfo{pages}{656--665}.
\newblock


\bibitem[\protect\citeauthoryear{Eldawy and Mokbel}{Eldawy and Mokbel}{}]%
        {eldawy2015spatialhadoop}
\bibfield{author}{\bibinfo{person}{Ahmed Eldawy} {and}
  \bibinfo{person}{Mohamed~F Mokbel}.}
\newblock \showarticletitle{SpatialHadoop: A MapReduce framework for spatial
  data}. In \bibinfo{booktitle}{{\em ICDE, year={2015}}}.
\newblock


\bibitem[\protect\citeauthoryear{Finkel and Bentley}{Finkel and
  Bentley}{1974}]%
        {finkel1974quad}
\bibfield{author}{\bibinfo{person}{Raphael~A. Finkel} {and}
  \bibinfo{person}{Jon~Louis Bentley}.} \bibinfo{year}{1974}\natexlab{}.
\newblock \showarticletitle{Quad trees a data structure for retrieval on
  composite keys}.
\newblock \bibinfo{journal}{{\em Acta informatica\/}} \bibinfo{volume}{4},
  \bibinfo{number}{1} (\bibinfo{year}{1974}), \bibinfo{pages}{1--9}.
\newblock


\bibitem[\protect\citeauthoryear{Grigni and Manne}{Grigni and Manne}{1996}]%
        {grigni1996complexity}
\bibfield{author}{\bibinfo{person}{Michelangelo Grigni} {and}
  \bibinfo{person}{Fredrik Manne}.} \bibinfo{year}{1996}\natexlab{}.
\newblock \showarticletitle{On the complexity of the generalized block
  distribution}.
\newblock In \bibinfo{booktitle}{{\em Parallel Algorithms for Irregularly
  Structured Problems}}. \bibinfo{publisher}{Springer},
  \bibinfo{pages}{319--326}.
\newblock


\bibitem[\protect\citeauthoryear{Guttman}{Guttman}{1984}]%
        {guttman1984r}
\bibfield{author}{\bibinfo{person}{Antonin Guttman}.}
  \bibinfo{year}{1984}\natexlab{}.
\newblock \bibinfo{booktitle}{{\em R-trees: a dynamic index structure for
  spatial searching}}. Vol.~\bibinfo{volume}{14}.
\newblock \bibinfo{publisher}{ACM}.
\newblock


\bibitem[\protect\citeauthoryear{Li, Lee, Zheng, Lee, Lee, and Wang}{Li
  et~al\mbox{.}}{2011}]%
        {li2011ir}
\bibfield{author}{\bibinfo{person}{Zhisheng Li}, \bibinfo{person}{Ken~CK Lee},
  \bibinfo{person}{Baihua Zheng}, \bibinfo{person}{Wang-Chien Lee},
  \bibinfo{person}{Dik Lee}, {and} \bibinfo{person}{Xufa Wang}.}
  \bibinfo{year}{2011}\natexlab{}.
\newblock \showarticletitle{Ir-tree: An efficient index for geographic document
  search}.
\newblock \bibinfo{journal}{{\em TKDE\/}} \bibinfo{volume}{23},
  \bibinfo{number}{4} (\bibinfo{year}{2011}), \bibinfo{pages}{585--599}.
\newblock


\bibitem[\protect\citeauthoryear{Ma, Zhang, and Meng}{Ma et~al\mbox{.}}{2013}]%
        {ma2013st}
\bibfield{author}{\bibinfo{person}{Youzhong Ma}, \bibinfo{person}{Yu Zhang},
  {and} \bibinfo{person}{Xiaofeng Meng}.} \bibinfo{year}{2013}\natexlab{}.
\newblock \showarticletitle{ST-HBase: a scalable data management system for
  massive geo-tagged objects}.
\newblock In \bibinfo{booktitle}{{\em Web-Age Information Management}}.
  \bibinfo{publisher}{Springer}.
\newblock


\bibitem[\protect\citeauthoryear{Mahmood, Aly, Qadah, Rezig, Daghistani,
  Madkour, Abdelhamid, Hassan, Aref, and Basalamah}{Mahmood
  et~al\mbox{.}}{2015}]%
        {mahmood2015tornado}
\bibfield{author}{\bibinfo{person}{Ahmed~R Mahmood}, \bibinfo{person}{Ahmed~M
  Aly}, \bibinfo{person}{Thamir Qadah}, \bibinfo{person}{El~Kindi Rezig},
  \bibinfo{person}{Anas Daghistani}, \bibinfo{person}{Amgad Madkour},
  \bibinfo{person}{Ahmed~S Abdelhamid}, \bibinfo{person}{Mohamed~S Hassan},
  \bibinfo{person}{Walid~G Aref}, {and} \bibinfo{person}{Saleh Basalamah}.}
  \bibinfo{year}{2015}\natexlab{}.
\newblock \showarticletitle{Tornado: A distributed spatio-textual stream
  processing system}.
\newblock \bibinfo{journal}{{\em PVLDB\/}} \bibinfo{volume}{8},
  \bibinfo{number}{12} (\bibinfo{year}{2015}), \bibinfo{pages}{2020--2023}.
\newblock


\bibitem[\protect\citeauthoryear{Mokbel and Aref}{Mokbel and Aref}{2008}]%
        {mokbel2008sole}
\bibfield{author}{\bibinfo{person}{Mohamed~F Mokbel} {and}
  \bibinfo{person}{Walid~G Aref}.} \bibinfo{year}{2008}\natexlab{}.
\newblock \showarticletitle{SOLE: scalable on-line execution of continuous
  queries on spatio-temporal data streams}.
\newblock \bibinfo{journal}{{\em The VLDB Journal\/}} \bibinfo{volume}{17},
  \bibinfo{number}{5} (\bibinfo{year}{2008}), \bibinfo{pages}{971--995}.
\newblock


\bibitem[\protect\citeauthoryear{Ooi, McDonell, and Sacks-Davis}{Ooi
  et~al\mbox{.}}{1987}]%
        {ooi1987spatial}
\bibfield{author}{\bibinfo{person}{Beng~Chin Ooi}, \bibinfo{person}{Ken~J
  McDonell}, {and} \bibinfo{person}{Ron Sacks-Davis}.}
  \bibinfo{year}{1987}\natexlab{}.
\newblock \showarticletitle{Spatial kd-tree: An indexing mechanism for spatial
  databases}. In \bibinfo{booktitle}{{\em IEEE COMPSAC}},
  Vol.~\bibinfo{volume}{87}. \bibinfo{pages}{85}.
\newblock


\bibitem[\protect\citeauthoryear{Samet}{Samet}{1990}]%
        {samet1990design}
\bibfield{author}{\bibinfo{person}{Hanan Samet}.}
  \bibinfo{year}{1990}\natexlab{}.
\newblock \bibinfo{booktitle}{{\em The design and analysis of spatial data
  structures}}. Vol.~\bibinfo{volume}{85}.
\newblock \bibinfo{publisher}{Addison-Wesley Reading, MA}.
\newblock


\bibitem[\protect\citeauthoryear{Toshniwal, Taneja, Shukla, Ramasamy, Patel,
  Kulkarni, Jackson, Gade, Fu, Donham, et~al\mbox{.}}{Toshniwal
  et~al\mbox{.}}{2014}]%
        {toshniwal2014storm}
\bibfield{author}{\bibinfo{person}{Ankit Toshniwal}, \bibinfo{person}{Siddarth
  Taneja}, \bibinfo{person}{Amit Shukla}, \bibinfo{person}{Karthik Ramasamy},
  \bibinfo{person}{Jignesh~M Patel}, \bibinfo{person}{Sanjeev Kulkarni},
  \bibinfo{person}{Jason Jackson}, \bibinfo{person}{Krishna Gade},
  \bibinfo{person}{Maosong Fu}, \bibinfo{person}{Jake Donham}, {and}
  \bibinfo{person}{others}.} \bibinfo{year}{2014}\natexlab{}.
\newblock \showarticletitle{Storm@ twitter}. In \bibinfo{booktitle}{{\em
  SIGMOD}}. ACM, \bibinfo{pages}{147--156}.
\newblock


\bibitem[\protect\citeauthoryear{Wang, Zhang, Zhang, Lin, and Wang}{Wang
  et~al\mbox{.}}{2015}]%
        {wang2015ap}
\bibfield{author}{\bibinfo{person}{Xiang Wang}, \bibinfo{person}{Ying Zhang},
  \bibinfo{person}{Wenjie Zhang}, \bibinfo{person}{Xuemin Lin}, {and}
  \bibinfo{person}{Wei Wang}.} \bibinfo{year}{2015}\natexlab{}.
\newblock \showarticletitle{Ap-tree: Efficiently support continuous
  spatial-keyword queries over stream}. In \bibinfo{booktitle}{{\em ICDE}}.
  \bibinfo{pages}{1107--1118}.
\newblock


\bibitem[\protect\citeauthoryear{Zaharia, Chowdhury, Franklin, Shenker, and
  Stoica}{Zaharia et~al\mbox{.}}{2010}]%
        {spark}
\bibfield{author}{\bibinfo{person}{Matei Zaharia}, \bibinfo{person}{Mosharaf
  Chowdhury}, \bibinfo{person}{Michael~J Franklin}, \bibinfo{person}{Scott
  Shenker}, {and} \bibinfo{person}{Ion Stoica}.}
  \bibinfo{year}{2010}\natexlab{}.
\newblock \bibinfo{title}{Spark: cluster computing with working sets.}
\newblock   (\bibinfo{year}{2010}).
\newblock


\bibitem[\protect\citeauthoryear{Zaharia, Das, Li, Hunter, Shenker, and
  Stoica}{Zaharia et~al\mbox{.}}{2013}]%
        {Sparkstreaming}
\bibfield{author}{\bibinfo{person}{Matei Zaharia}, \bibinfo{person}{Tathagata
  Das}, \bibinfo{person}{Haoyuan Li}, \bibinfo{person}{Timothy Hunter},
  \bibinfo{person}{Scott Shenker}, {and} \bibinfo{person}{Ion Stoica}.}
  \bibinfo{year}{2013}\natexlab{}.
\newblock \bibinfo{title}{Discretized streams: Fault-tolerant streaming
  computation at scale}.
\newblock   (\bibinfo{year}{2013}).
\newblock


\bibitem[\protect\citeauthoryear{Zhang, Ma, and Meng}{Zhang
  et~al\mbox{.}}{2014}]%
        {zhang2014efficient}
\bibfield{author}{\bibinfo{person}{Yu Zhang}, \bibinfo{person}{Youzhong Ma},
  {and} \bibinfo{person}{Xiaofeng Meng}.} \bibinfo{year}{2014}\natexlab{}.
\newblock \showarticletitle{Efficient Spatio-textual Similarity Join Using
  MapReduce}. In \bibinfo{booktitle}{{\em IAT}}, Vol.~\bibinfo{volume}{1}.
  \bibinfo{pages}{52--59}.
\newblock


\bibitem[\protect\citeauthoryear{Zobel and Moffat}{Zobel and Moffat}{2006}]%
        {zobel2006inverted}
\bibfield{author}{\bibinfo{person}{Justin Zobel} {and}
  \bibinfo{person}{Alistair Moffat}.} \bibinfo{year}{2006}\natexlab{}.
\newblock \showarticletitle{Inverted files for text search engines}.
\newblock \bibinfo{journal}{{\em ACM computing surveys (CSUR)\/}}
  \bibinfo{volume}{38}, \bibinfo{number}{2} (\bibinfo{year}{2006}),
  \bibinfo{pages}{6}.
\newblock


\end{thebibliography}

\end{document}